\documentclass[a4paper,reqno]{amsart}
\usepackage[text={5.2in,8in},centering]{geometry}

\usepackage[usenames,dvipsnames,svgnames,table]{xcolor}
\definecolor{citegreen}{rgb}{0.00,0.70,0.30}
\usepackage[colorlinks=true,
            linkcolor=blue,
            urlcolor=blue,
            citecolor=blue]{hyperref}

\usepackage{amsrefs}
\usepackage{amssymb}

\usepackage{amsmath}
\usepackage{amsthm}

\usepackage{mathrsfs}

\usepackage{appendix}

\usepackage{xspace}

\usepackage{enumerate}
\usepackage{graphicx}

\numberwithin{equation}{section}
\theoremstyle{plain}
\newtheorem{theorem}{Theorem}
\newtheorem{lemma}{Lemma}[section]

\newtheorem{prop}[lemma]{Proposition}
\newtheorem{cor}[lemma]{Corollary}

\theoremstyle{remark}
\newtheorem{remark}[theorem]{Remark}

\newtheorem*{quest*}{Question}
\newtheorem*{remark*}{Remark}
\theoremstyle{definition}
\newtheorem{definition}{Definition}[section]

\newtheorem*{notation*}{Notation}
\newtheorem*{notations*}{Notations}
\providecommand{\B}{\mathbf}

\providecommand{\D}{\mathbb}

\newcommand{\ee}{\mathrm{e}}

\newcommand{\ii}{\mathrm{i}}

\providecommand{\esm}[1]{\D{E}\left[ #1 \right]}

\DeclareMathOperator{\dist}{dist}
\DeclareMathOperator{\diam}{diam}
\DeclareMathOperator{\supp}{supp}
\DeclareMathOperator{\card}{card}

\DeclareMathOperator*{\essup}{ess\,sup}

\DeclareMathOperator{\one}{\mathbf{1}}

\definecolor{redd}{rgb}{0.95,0.2,0.2}
\definecolor{gris}{rgb}{0.9,0.9,0.9}
\definecolor{greenn}{rgb}{0.1,0.6,0.2}

\definecolor{cmgray}{rgb}{0.7,0.7,0.7}

\definecolor{cmblue}{rgb}{0.2,0.5,0.8}

\DeclareMathAlphabet{\mathpzc}{OT1}{pzc}{m}{it}

\def\mes{{\mathrm mes}}

\def\be{\begin{equation}}
\def\ee{\end{equation}}
\def\ba{\begin{array}{l}}
\def\ea{\end{array}}
\def\bal{\begin{aligned}}
\def\eal{\end{aligned}}

\def\ble{\begin{lemma}}
\def\ele{\end{lemma}}

\def\bthm{\begin{theorem}}
\def\ethm{\end{theorem}}

\def\bco{\begin{cor}}
\def\eco{\end{cor}}

\def\bpr{\begin{prop}}
\def\epr{\end{prop}}

\def\bre{\begin{remark}}
\def\ere{\end{remark}}

\def\btm{\begin{theorem}}
\def\etm{\end{theorem}}

\def\bde{\begin{definition}}
\def\ede{\end{definition}}

\def\eu{{\mathrm{e}}}

\def\ffi{\varphi}

\def\BLam{\boldsymbol{\Lambda}}

\def\BDelta{\boldsymbol{\Delta}}

\def\bcubeout{\boldsymbol{\Lambda}^{\text{out}}}

\def\half{\frac{1}{2}}
\def\shalf{{\textstyle{\frac{1}{2}}}}
\def\third{\frac{1}{3}}

\def\quart{\frac{1}{4}}

\def\EdmS{$(E,\delta,m)$-{\rm S}\xspace}

\def\eps{\epsilon}
\def\lam{\lambda}

\def\pt{\partial}
\def\Const{\mathrm{Const\,}}

\def\pr#1{\mathbb{P}\left\{ #1 \right\}}
\def\esm#1{\D{E}\left[\, #1\, \right]}
\def\Bigesm#1{\D{E}\Big[\, #1\, \Big]}
\def\bigesm#1{\D{E}\big[\,#1\,\big]}

\newcommand{\vertii}[1]{{\big\vert\kern-0.25ex\big\vert #1
    \big\vert\kern-0.25ex\big\vert\kern-0.25ex}}

\newcommand{\vertiii}[1]{{\big\vert\kern-0.25ex\big\vert\kern-0.25ex\big\vert #1
    \big\vert\kern-0.25ex\big\vert\kern-0.25ex\big\vert}}

\newcommand{\dnorm}[1]{{\big\| #1 \big\|^{\curlywedge}}}

\def\tabup{\rule{0pt}{4ex}}
\def\tabdown{\rule[-3ex]{0pt}{0pt}}

\def\BC{\mathbf{C}}
\def\bcell{\mathbf{C}}

\def\bcubkoneu{\bcube_{L_{k+1}}(\Bu)}

\def\Cgeom{C^{\rm GRI}}

\def\lam{\lambda}
\def\om{\omega}
\def\th{\theta}
\def\ffi{\varphi}

\def\Lam{\Lambda}
\def\Om{\Omega}

\def\Sigmani{\Sigma^{\rm ni}}

\def\vempty{\varnothing}

\def\EdS{{\rm$(E,\delta,m_N)$-S}\xspace}
\def\EdNS{$(E,\delta,m_N)$-NS\xspace}

\def\ER{{\rm$(E,\beta)$-R}\xspace}
\def\ENR{$(E,\beta)$-NR\xspace}

\def\Egood{{\rm$(E,\delta,m_N)$-good}\xspace}
\def\Ebad{$(E,\delta,m_N)$-bad\xspace}

\def\RCM{\textsf{(RCM)}\xspace}

\def\Vone{\textsf{(V)}\xspace}

\def\Uone{\textsf{(U)}\xspace}

\def\Unif{{\mathrm{Unif}}}

\def\bcube{{\boldsymbol{\Lambda}}}
\def\bcubeN{\boldsymbol{\Lambda}^{N}}

\def\cube{\Lambda}

\def\tbcube{\widetilde{\boldsymbol{\Lambda}}}
\def\tbball{\widetilde{\mathbf{B}}}
\def\ball{\mathrm{B}}

\def\fS{\mathfrak{S}}
\def\fF{\mathfrak{F}}

\def\BA{\mathbf{A}}
\def\BB{\mathbf{B}}

\def\BF{\mathbf{F}}

\def\BG{\mathbf{G}}
\def\BGni{\mathbf{G}^{\mathrm{ni}}}

\def\BH{\mathbf{H}}
\def\BHni{\mathbf{H}^{\mathrm{ni}}}

\def\pbar{\overline{p}}

\def\BP{\mathbf{P}}
\def\BU{\mathbf{U}}
\def\BV{\mathbf{V}}

\def\ibA{\boldsymbol{A}}
\def\ibB{\boldsymbol{B}}

\def\BPsi{\boldsymbol{\Psi}}

\def\bcA{\boldsymbol{\mathcal{A}}}

\def\bcX{\boldsymbol{\mathcal{X}}}

\def\bcZ{\boldsymbol{\mathcal{Z}}}

\def\bcZN{\boldsymbol{\mathcal{Z}}^{N}}

\def\SS#1{{\textsf{S}$\left(#1\right)$}}

\def\bcubeLk{\BLam_{L_{k}}}
\def\bcubeLkone{\BLam_{L_{k+1}}}

\def\csA{\mathscr{A}}

\def\csE{\mathscr{E}}

\def\csR{\mathscr{R}}
\def\csS{\mathscr{S}}

\def\cB{\mathcal{B}}
\def\cE{\mathcal{E}}

\def\cG{\mathcal{G}}
\def\cJ{\mathcal{J}}
\def\cL{\mathcal{L}}
\def\cM{\mathcal{M}}

\def\cS{\mathcal{S}}
\def\cbS{\boldsymbol{\mathcal{S}}}
\def\cV{\mathcal{V}}
\def\cZ{\mathcal{Z}}

\def\bball{\mathbf{B}}
\def\bballN{\mathbf{B}^{(N)}}

\def\Ba{\mathbf{a}}
\def\Bb{\mathbf{b}}
\def\Bx{\mathbf{x}}
\def\By{\mathbf{y}}
\def\Bz{\mathbf{z}}
\def\Bu{\mathbf{u}}
\def\Bv{\mathbf{v}}

\def\rc{\mathrm{c}}
\def\rd{\mathrm{d}}

\def\rr{\mathrm{r}}
\def\rw{\mathrm{w}}

\def\rC{\mathrm{C}}

\def\rM{\mathrm{M}}
\def\rP{\mathrm{P}}
\def\rQ{\mathrm{Q}}
\def\rS{\mathrm{S}}

\def\DP{\mathbb{P}}
\def\DR{\mathbb{R}}
\def\DZ{\mathbb{Z}}
\def\DN{\mathbb{N}}

\begin{document}
\title[Localization in a continuous multi-particle Anderson model]
{Efficient localization bounds\\in a continuous multi-particle Anderson model\\with long-range interaction}
\author{Victor Chulaevsky$^1$}
\date{}
\begin{abstract}
We establish strong dynamical localization for a class of multi-particle Anderson models in a Euclidean space
with an alloy-type random potential and a sub-exponentially decaying interaction of infinite range.
For the first time in the mathematical literature, the uniform decay bounds on the eigenfunction correlators at low energies
are proved, in the multi-particle continuous configuration space, in the norm-distance and not in the
Hausdorff pseudo-metric.
\end{abstract}
\maketitle
\section{Introduction} \label{sec:intro}
\subsection{The model}

We study a multi-particle Anderson model in $\DR^d$ with long-range interaction
and subject to an external random potential of the so-called alloy type.
The Hamiltonian $\B{H}$ $\left(=\B{H}^{(N)}(\omega)\right)$ is a random Schr\"{o}dinger operator of the form
\begin{equation}\label{eq:def.H}
\B{H}=-\frac{1}{2}\B{\Delta}+\B{U}(\Bx)+\B{V}(\omega;\Bx)
\end{equation}
acting in $L^2\big( (\DR^{d})^N\big)$. To stress the dependence on the number of particles, $n\ge 1$, omitting a less
important parameter $d $ ($=$ the dimension of the $1$-particle configuration space), we denote
$$
\bcX^N := \big(\DR^d\big)^N,  \;\; \bcZ^N := \big(\DZ^d\big)^N \hookrightarrow \big(\DR^d\big)^N, \;\; N\ge 1.
$$
The points $\Bx=(x_1,\ldots,x_N)\in \bcX^N$ represent the positions of the $N$ quantum particles evolving simultaneously
in the physical space $\DR^d$.
In \eqref{eq:def.H},
$\BDelta$ stands for the Laplacian in $(\DR^d)^N$ (or, equivalently, in $\DR^{Nd}$). The interaction energy
operator $\BU$ acts as multiplication by a function $\Bx \mapsto \BU(\Bx)$. Finally, the potential energy $\BV(\omega;\Bx)$
(unrelated to the inter-particle interaction)
is the operator of multiplication by a function
\begin{equation}\label{eq:def.V}
\Bx\mapsto V(x_1;\omega)+ \cdots + V(x_N;\omega),
\end{equation}
where $x\in\DR^d\mapsto V(x;\omega)$ is a random external field potential assumed to be of the form
\begin{equation}\label{eq:defalloy}
V(x;\omega)=\sum_{a\in\D{Z}^d}\cV_a(\omega)\, \ffi(x-a).
\end{equation}
Here and below $\cV_a$, $a\in\D{Z}^d \hookrightarrow \DR^d$, are IID (independent and identically
distributed) real random variables on some probability space $(\Omega,\fF,\DP)$ and
$\ffi:\DR^d\to\DR$ is usually referred to as a scatterer (or ``bump'') function.

More precise assumptions will be specified below.

\subsection{The motivation and comparison with the existing results}

The single-particle localization theory, describing non-interacting quantum particles in a random environment,
was initiated by P. W. Anderson in his seminal paper \cite{A58}, which he affectionately called the "\emph{Nobel Prize One}".
The first rigorous mathematical results on random Anderson-type Hamiltonians were obtained by Goldsheid et al. \cite{GMP77}
(in $\DR^1$), Kunz and Souillard \cite{KS80} (in $\DZ^1$), and then in multi-dimensional lattice models by Fr\"{o}hlich et al.
\cite{FS83,FMSS85}, with the help of the Multi-Scale Analysis (MSA). An alternative approach was developed by Aizenman and Molchanov
\cite{AM93}, for the lattice models; the original technique was later substantially generalized
in a series of deep works bearing a distinctive mark of Michael Aizenman's enthusiasm; cf., e.g., \cite{Ai94,ASFH01,AENSS06}.

Martinelli and Holden \cite{MH84} extended the MSA to the continuous models, i.e., to the random Hamiltonians in $L^2(\DR^d)$.

In all the above mentioned papers, as well as in a number of other physical and mathematical works, the quantum particles
were considered non-interacting; in the physical context, this is of course a conscious approximation, made already by
Anderson \cite{A58} who did not hide his concerns about the possible effects of the interaction on the localization phenomena.

An outburst of new results, both in theoretical and in mathematical physics, occurred in 2005-2008 (cf. \cite{BAA06}, \cite{GorMP05},
\cite{CS08,CS09a,CS09b,AW09a}; some preprints appeared earlier).
As usual, the physical works provided stronger statements, viz. the stability of localization phenomena
under sufficiently weak interaction in physicaly realistic systems, with $N \sim \rho |\Lam|$ particles in a bounded
but macroscopical large domain $\Lam\subset\DR^d$. We stress that $\Lam$ is indeed of \emph{finite}, albeit possibly large size.
It goes without saying that for all imaginable applications the size of $\Lam$ is bounded by (or is of order of) that of our little planet,
and \emph{not actually infinite}. The first mathematical works  considered a fixed number $N>1$ of particles
in an infinite configuration space: $\DZ^d$ in \cite{CS08,CS09a,CS09b,AW09a}, and later $\DR^d$ (cf. \cite{CBS11},
\cite{Sab13}, \cite{KN13b}, \cite{FW14}).

While it does not come as a big surprise that only a finite, and fixed, number of particles were allowed in the
first attempts to build rigorous theory of Anderson localization in systems with nontrivial interaction, it is more surprising
that the results on complete spectral and strong dynamical localization proved in \cite{CS09b,AW09a} \emph{required} the configuration space
to be \emph{actually infinite}. More precisely, localization (viz., uniform bounds on the eigenfunction correlators, or the
rate of spread of an initially localized wavepacket $\psi$ under the Hamiltonian dynamics $\eu^{-\ii t \BH}$) could not be established
in arbitrarily large yet bounded domain $\Lam$ of the physical configuration space.

This is quite opposite to the usual situation where a finite-volume analysis is only a prelude for the rigorous study of
the object inspiring mathematicians -- an actually infinite system. If the results of \cite{CS09b,AW09a} (or their continuous-space
counterparts \cite{CBS11,KN13b,FW14}) were to be applied to the physical models, they would be valid only if our Universe
were found to be infinite. Otherwise, one could not be able to rule out some tunneling processes which might result in
transfer of particles over arbitrarily large distances.

On the technical level, the bottleneck of previously available rigorous results on the $N$-particle localization
(starting with $N=3$) is the eigenvalue concentration (EVC) estimate, which is analogous to, but more sophisticated than,
its well-known counterpart going back to Wegner \cite{W81}.  Despite significant differences between the techniques
of \cite{CS09b} and \cite{AW09a},
both approaches faced essentially the same problem, and both  gave rise to the decay bounds on the eigenfunctions
(EFs) and eigenfunction correlators (EFCs) expressed in terms of the so-called Hausdorff (pseudo-)distance but not the norm-distance.

\vskip1mm

Now we turn to the goals and results of the present work.

\vskip1mm

Following the approach to the multi-particle EVC bounds presented originally in \cite{C10,C12b} and recently extended in \cite{C13a},
we aim to improve the EVC estimates required for the multi-particle MSA (MPMSA) and achieve more efficient decay $N$-particle
localization bounds for $N\ge 3$ particles in a Euclidean space $\DR^d$, $d\ge 1$. For the lattice systems, such a task was performed
in our recent work \cite{CS14}.

\vskip1mm

$\blacklozenge$ The main novelty of the present work is the first rigorous proof (for $N\ge 3$) of uniform decay
(which we show to be at least sub-exponential)
of the eigenfunction correlators with respect
to a genuine norm-distance in a multi-particle alloy model in $\DR^d$ or in any bounded regular sub-domain thereof.
In accordance with the above discussion, the phenomenon of Anderson localization is thus firmly established in disordered systems
with fixed number of quantum particles in a physically realistic geometrical setting.
For the moment, this result is proved for
a particular class of alloy potentials, which we call \emph{flat tiling} alloys.

$\blacklozenge$  Further, compared to the paper by Klein and Nguyen \cite{KN13b}, who made a significant step in the scaling analysis
of the continuous interactive multi-particle Anderson models,
the main improvement is relaxing the condition of finite-range interaction
to a sub-exponential decay of the interaction potential.

Surprisingly, in the models with sub-exponential decay of the interaction potential,
the MPMSA approach provides a
qualitatively stronger decay rate (a genuine exponential one) of the localized EFs than the new variant of the MPFMM
developed by Fauser and Warzel \cite{FW14}. The decay rate of the EFs derived from the analysis of the
EF correlators, naturally, cannot be stronger than that of the EFCs, while the (MP)MSA scheme is free from this
logical dependence. We postpone the proof of exponential decay of the localized EFs to a forthcoming paper, but
note that this can be achieved by way of straightforward, ``algotithmic'' adaptations of some known results:
\par\vskip1mm\noindent
$\bullet$ First, one can follow the bootstrap MPMSA strategy developed by Klein and Nguyen,
simply replacing the two-volume EVC
bound \cite{KN13b}*{Corollary 2.3}, applicable to the Hausdorff--distant pairs of cubes,
by the one proved in the present paper (cf. Theorem \ref{thm:W2}) and applicable to the norm-distant pairs.
A careful reading of Ref.~\cite{KN13b} evidences that the extension of their techniques to sub-exponentially
decaying interactions can be carried out with the help of the perturbation argument from Lemma \ref{lem:WITRONS.subexp}
(complementing the analysis in \cite{KN13b}*{Sect.~5.4.2}). We also point out here that the impact of the choice of the Hausdorff- or norm-distance
in the $N$-particle configuration space on the qualitative result regarding the exponential decay of the \emph{eigenfunctions},
particularly in the infinite configuration space $(\DR^d)^N$,
is much less pronounced than for the decay of the EF \emph{correlators}.

\par\vskip1mm\noindent
$\bullet$ Alternatively, one can employ the approach to the derivation of variable-energy MSA estimates from their fixed-energy counterparts,
described in \cite{C14a} (cf. Theorem 4) or, more precisely, an extension thereof to self-adjoint operators with compact resolvent
(Ref.~\cite{C14a} treated lattice systems, where the finite-volume Hamiltonians are finite-dimensional).
We plan to discuss such an
adaptation
in a separate paper.

\vskip2mm
It is to be emphasized that the recent result by Fauser and Warzel \cite{FW14} on exponential decay of the EFCs, for exponentially
decaying interactions, remains the strongest one \emph{among those proved in terms of Hausdorff distance}
in a continuous space, hence, in an \emph{actually infinite} configuration space.
Due to some well-known limitations of the Multi-Scale Analysis (single- or multi-particle), proofs of exponential
strong dynamical localization are still beyond the MSA' reach. On the other hand, recall that the technique developed by Klein and
Nguyen \cite{KN13b},
based on the Quntitative Unique Continuation Principle (QUCP), made unnecessary the complete covering condition for the alloy potential,
used both in \cite{CBS11} and in \cite{FW14}. This makes the class of models considered in \cite{KN13b} the most general one,
at the time of writing these lines.

It seems appropriate to attract the readers' attention to an interesting fact: while one of the most striking differences
between the MSA and the FMM, in the single-particle localization theory, is that the latter employs a "mono-scale" technology, in
the world of multi-particle systems both approaches -- MSA and FMM -- finally settle on the common ground of
\emph{multi-scale} geometrical induction.

\vskip2mm
Except for the new EVC bounds, the main strategy of the proofs in the present paper is an improved and simplified variant of the
MPMSA from \cite{CBS11}, with elements of the bootstrap MSA.
To be more precise, we do not actually make a \emph{bootstrap}, but rather carry out two logically independent scaling analyses,
analogous (but not identical) to two of the four phases of the bootstrap MSA (cf. \cite{GK01,KN13a,KN13b}).
This simplification has however
its drawbacks. Below we comment, where appropriate, on
the important advantages of the full-fledged bootstrap analysis; the task of performing such analysis is beyond the scope of the present paper,
which we intentionally keep relatively short.

In a forthcoming work, we plan to address the multi-particle alloy models with lower
regularity of the marginal disorder distribution, and with more general structure of the scatterer functions, without the complete covering (let alone
flat tiling) condition. In such models, a minor modification of our scheme gives rise to the EFC decay estimates in the Hausdorff distance.


\subsection{Basic geometric objects and  notations}

Throughout this paper, we will fix an integer $N^* \geq 2$ and work in Euclidean spaces of the
form $(\DR^{d})^N\cong\DR^{Nd}$, $1\le N \le N^*$. A configuration of $N\ge 1$ distinguishable particles
in $\DR^d$ is represented by (and in our paper, identified with) a vector $\Bx=(x_1, \ldots, x_N)\in(\DR^d)^N$,
where $x_j$ is the position of the $j$-th particle.
More generally, boldface notations are reserved for "multi-particle" objects (Hamiltonians, resolvents, cubes, etc.).

All Euclidean spaces will be endowed with the max-norm denoted by $|\,\cdot\,|$. We will consider
$Nd$-dimensional cubes of integer edge length in
$(\DZ^d)^N$ centered at lattice points $\Bu\in(\DZ^d)^N \hookrightarrow (\DR^d)^N$ and with edges parallel to the co-ordinate axes.
The open cube of edge length $2L+1$ centered at $\Bu$ is denoted by $\bcube_L(\Bu)$; in the max-norm
it represents the open ball of radius $L+\half$ centered at $\Bu$:
\begin{equation}\label{eq:BLam}
\bcube_L(\B{u}) =\{\Bx\in\bcX^N: \;|\Bx - \Bu| < L + \shalf\}.
\end{equation}
The lattice counterpart for $\bcube_L(\Bu)$ is denoted by $\bball_L(\Bu)$:
$$
\bball_L(\Bu) = \bcube_L(\Bu) \cap (\DZ^d)^N; \quad \Bu\in(\DZ^d)^N.
$$
We also consider ``cells'' -- closed cubes of diameter $1$ centered at lattice points $\Bu\in\bcZN$:
$$
\BC(\Bu) = \{\By\in\bcX^n: \; |\By - \Bu| \le \shalf \}
$$
The union of all cells $\BC(\Bu)$, $\Bu\in\bcZN$, covers the entire Euclidean space $\bcX^N$.
Moreover, for any $\Bu\in\bcZ^N$, denoting by $\overline{\ibA}$ the closure of the set $\ibA\subset\bcX^N$, we have
$$
\overline{\bcube}_L(\Bu) = \bigcup_{ \By \in \bball_L(\Bu)} \bcell(\By).
$$

The diameters appearing in our formulae are relative to the max-norm;
the cardinality of various sets $A$ (usually finite) will be denoted by $|A|$. We have
$$
\diam \bcube_L(\Bu) = 2L+1, \diam \bball_L(\Bu) = 2L, \;\; | \bball_L(\Bu)| = (2L+1)^{nd} \le (3L)^{nd}.
$$

The indicator function of a set $A$ is denoted in general by $\one_A$, but for the indicators of the cells
we use a shorter notation, $\chi_\Bx := \one_{\bcell(\Bx)}$.

\subsection{Symmetrized norm-distance and the Hausdorff metric}

A norm-distance in the $N$-particle configuration space is not well-adapted to the decay estimates of the eigenfunctions
and of their correlators. Indeed, if the interaction potential $\BU$ is permutation-symmetric (and the external
random potential is always so), then the entire Hamiltonian $\BH^{(N)}$ commutes with the symmetric group $\fS_N$ acting
by permutations of the particle positions. Thus the Hilbert space $L^2(\bcX^N)$ can be decomposed in the direct sum
of $\BH^{(N)}$-invariant subspaces, including, for example, that of the symmetric functions taking identical values along
any orbit of the symmetry group $\fS_N$. The points of such orbits can be arbitrarily distant from each other,
which makes impossible any uniform decay bound.

More to the point, the physical systems are composed of indistinguishable particles,
so the permutations of the particle positions give rise to equivalent configurations.

For these reasons, the symmetrized norm-distance in the $N$-particle configuration space is much more
natural, even in a situation where, as in the present paper, the particle are considered distinguishable.
The formal definition is as follows:
$$
\rd^{(N)}_S( \Bx, \By) := \min_{\pi\in\fS_N} \| \pi(\Bx) - \By \|_\infty,
$$
(we choose the max-norm $\|\Bx\|_\infty = \max_j |x_j|$)
where the elements of the symmetric group $\pi\in\fS_N$ act on $\Bx = (x_1, \ldots, x_N)$ by permutations of the coordinates $x_j$.

Recall the definition of the Hausdorff distance $\rd_H$ between two subsets $X,Y$ of a metric space $(\cM, \rd(\cdot))$:
$$
\rd_H(X, Y) = \max\Big[\; \sup_{x\in X} \rd(x, Y), \sup_{y\in Y} \rd(y, X)  \; \Big] .
$$
This notion does not apply directly to the configurations $\Bx\in\bcX^N$; however, an important characteristics of
$\Bx = (x_1, \ldots, x_N)$ is its "projection" $\Pi \Bx = \{x_1, \ldots, x_N\}$. In the case of indistinguishable Fermi-particles,
$\Pi\Bx$ \emph{is} the configuration.

It was discovered in \cite{CS09b,AW09a} that the decay bounds on the Green function, eigenfunctions and eigenfunction correlators
were simpler to obtain with respect to the Hausdorff distance $\rd^{(N)}_H(\Bx, \By) := \rd_H(\Pi\Bx, \Pi \By)$ (for $\Bx,\By\in\bcX^N$)
than in terms of the symmetrized norm-distance.

The following simple example illustrates the difference between $\rd^{(N)}_S$ and $\rd^{(N)}_H$,
for $N\ge 3$.
With $N=3$, $d=1$, let $\Ba = (0,0,R)$ and $\Bb = (0,R,R)$. Then
$\rd^{(3)}(\bcube_L(\Ba), \bcube_L(\Bb)) \to+\infty$ as $|R|\to \infty$, but
$\rd^{(N)}_H(\bcube_L(\Ba), \bcube_L(\Bb)) \equiv 0$.

In physical terms, $\Bb$ is obtained from $\Ba$ by
transferring\footnote{Such a transfer can be "partial", i.e., leaving at least one particle at each of the two distant loci $0$ and $R$,
for $N\ge 3$, while for $N=2$ a similar transfer must be "complete" in one of the two directions.}
one of the particles from $0$ to a distant location $R$.
If one has to study localization in a finite domain $[0,R]$, then the tunneling between the configurations like
$\Ba$ and $\Bb$ can (or might) ruin
the decay of EFs and EFCs over the distances comparable with the size of the domain.

Such a situation is impossible for $N=2$, since $\rd^{(2)}_H(\cdot\,,\cdot)$ is equivalent to $\rd^{(2)}_S(\cdot\,,\cdot)$.

\subsection{Interaction potential}

We assume the following:

\Uone $\BU$ is generated by a $2$-body potential $U^{(2)}:\DR_+\to \DR_+$, viz.
$$
\BU(\Bx) = \sum_{1\leq i < j\le N  }U^{(2)}(|x_i - x_j|),
$$
where
\be
0 \le U^{(2)}(r) \le C_U \eu^{-r^\zeta} ,
\ee
for some  $\zeta>0$, $C_U\in(0,\infty)$.

It does not make much sense to consider separately $\zeta>1$, for the key parameters measuring the decay of the EFCs
depend in fact upon the quantity $\min(1,\zeta)$.

\subsection{External random potential}
\label{ssec:cond.on.V}

We assume the following conditions to be fulfilled.

\Vone
The external random potential is of alloy type,
\be\label{eq:alloy}
V(x;\om) = \sum_{a\in\cZ} \cV(a;\om) \ffi_a(x - a),
\ee
where $\cV:\cZ\times\Om\to\DR$ is an IID random field on the lattice $\cZ = \DZ^d \hookrightarrow \DR^d$.

The scatterer (a.k.a. bump) functions $\ffi_a$ have the following property which we call \emph{\textbf{flat tiling}}:
$\diam \supp\, \ffi_a \le \rr_1 < \infty$ and
\be\label{eq:covering.condition}
\sum_{a\in\cZ} \ffi_a \equiv \rC_\ffi \one >0, \;\; \rC_\ffi \in(0,+\infty).
\ee

(\emph{In fact, our methods apply also to non-identical scatterers with flat tiling.})

The common marginal probability distribution of the IID scatterers amplitudes $\cV(\cdot;\om)$
admits a probability density $p_V$, which is compactly supported, with
$\supp\, p_V$ $ = [0, c_V]$, $c_V>0$, and $p_V$ is strictly positive, bounded and has bounded derivative
in the open interval $(0,c_V)$:
\be\label{eq:cond.pV}
\forall\, t\in (0,c_V)\qquad
\begin{cases} 0 < p_* \le p_V(t) \le p^* < +\infty \\ \qquad\qquad\, p'_V(t) \le C^* < +\infty
\end{cases}
\ee

Probably, the most natural example is where $\ffi = \one_{\overline{\cube}_{1/2}(0)}$, so that
$\sum_{a\in\cZ} \ffi(a)\equiv 1$.
Such a form of alloy was considered by Kotani and Simon \cite{KS87}, in the single-particle setting.
However, flat tiling is achieved
also for $\ffi = \one_{\overline{\cube}_{\ell/2}(0)}$, $\DN\ni\ell \ge 1$.
As to the scatterers' amplitudes, one can simply take the uniform probability distribution
$\Unif([0,1])$, where $c_V= C^* =p_* = p^*=1$.

For brevity, we assume $\rC_\ffi=1$; this is inessential for the validity of the main results.

\subsection {Main result}

Below we use the standard "bra--ket" notations $\langle \phi | H | \psi\rangle = (\phi, H \psi)$.
\begin{theorem}\label{thm:main}
Assume the conditions {\rm\Vone} and {\rm\Uone} and fix an integer $N^* \ge 2$.
There exist $\kappa=\kappa(\zeta,N^*) \in (0,\zeta)$, $\nu>0$ with the following property.
\par\vskip1mm
For all $N\in[1,N^*]$ and some nonrandom constant $C$, for all $\Bx,\By\in\bcX^N$ with $R:=\rd_S(\Bx,\By) \ge 1$,
and for any regular domain $\BLam\subseteq\bcX^N$ (bounded or not)
such that $\BLam\supset \bcube_{R/2}(\Bx) \cup \bcube_{R/2}(\By)$
\be\label{eq:thm.Main.DL.02}
\Bigesm{ \sup_{t\in\DR} \big| \langle \one_{\By}\, | \,  P_{I^*}\big(\BH^{(N)}_{\BLam}\big) \, \eu^{-\ii t \BH^{(N)}_{\BLam} } \, | \, \one_{\Bx}\rangle\big| }
\le C\, \eu^{ - \nu \left(\rd_{\rS}(\Bx,\By)\right)^{\kappa} }.
\ee
\end{theorem}

\section{EVC bounds}

\subsection{Regularity of the Conditional Mean (RCM)}

The key property of the probability distribution of the random scatterers in the flat tiling model,
resulting in efficient EVC bounds and ultimately, in norm-bounds on the decay of EFCs, can be formulated
for a random field $\cV:\cZ\times\Om\to\DR$ on a countable set $\cZ$ and relative to some probability space
$(\Om,\fF,\DP)$. Formally speaking, it does not presume independence or any explicit decay of correlations
of the random field in question.

Introduce the following notation. Given a finite set $Q\subset\cZ$, we set
$\xi_Q(\om) := |Q|^{-1}\sum_{x\in Q} V(x;\om)$ (the sample, or empirical, mean) and $\eta_x(\om) = V(x;\om) - \xi_Q(\om)$
for $x\in Q$ (the "fluctuations).

\vskip1mm

\RCM:
\emph{ Given a random field $\cV:\cZ\times\Om\to\DR$ on a countable set $\cZ$,
there exist constants $C', C'', A', A'', \th', \th''\in(0,+\infty)$ such that
for any finite subset $Q\subset\cZ$,
the (random) continuity modulus $\nu_Q(\cdot)$
of the conditional distribution  function
$F_\xi( \cdot \,| \fF_{Q})$ of the sample mean $\xi_Q$, defined by
\be\label{eq:def.nuQ}
\nu_Q(s; \om) :=  \sup_{t\in \DR} \; \essup \, \; \big|F_\xi(t+s\,| \fF_{Q}) - F_\xi(t\,| \fF_{Q}) \big|,
\ee
satisfies for all $s\in(0,1)$
\begin{equation}\label{eq:CMxi}
 \;\;
\pr{ \nu_Q(s; \om) \ge C' |Q|^{A'} s^{\th'} }
\le C''\, |Q|^{A''} s^{\th''}.
\end{equation}
} 

\bre
In fact, the explicit conditions \eqref{eq:cond.pV} on the marginal probability distribution of the scatterers
can be replaced (at least, for the IID scatterers' amplitudes) by \RCM, so the main result of the paper
remains valid under this more general hypothesis.
\ere

The condition \RCM is obviously fulfilled
for an IID Gaussian field, e.g., with zero mean and a unit variance; in this case
the sample mean
is independent of the fluctuations $\eta_\bullet$ and has a normal distribution with
variance $\sigma^2 = |Q|^{-1}$. An elementary argument (cf. \cite{C13a})
shows that \RCM also holds for an IID random field with a
uniform marginal distribution.
Moreover, using standard approximation techniques, one can prove the following result:
%
\bpr[Cf. \cite{C13a}*{Theorem 6}]\label{prop:RCM}
If a random field $\cV: \cZ\times \Om \to\DR$ obeys {\rm\Vone},
then it satisfies the property {\rm\RCM} with
$$
C' = 1, \, A' = 0, \, \th' = 2/3,
\;\; C'' = (4 C_p \pbar)^2,\, A'' =2, \,\th'' = 2/3.
$$
\epr

It is readily seen that a random field $\cV$ on $\cZ$, satisfying \RCM, can be decomposed
on any finite subset $Q\subset\cZ$ into the sum of a constant random field
$(\om,x)\mapsto \xi_Q \one_Q(x)$ and a "fluctuation" random field $\eta_x(\om)$, in such a way
that even with $\eta_x(\cdot)$ fixed by conditioning, the random constant still has
a sufficiently regular (conditional) probability distribution. In the context of random Anderson
Hamiltonians on $\cZ$ (which has to be endowed with the graph structure in this case),
the latter constant field acts in a very simple way on all EVs, and this results
in an elementary EVC bound. The reverse of the medal
is a non-optimal volume dependence, which, fortunately, never posed any problem for the
localization analysis.

The fact that the sample mean $\xi_Q$ modulates a \emph{constant} is crucial for our
proofs, and this is precisely the reason we assume the (alas, very restrictive) flat tiling condition.

\subsection{Bounds for the flat tiling alloy model}

We start with the one-volume EVC bound, which is quite similar in form to the celebrated Wegner estimate
\cite{W81}. The flat tiling alloy model is a particular case of the more general one, studied by
Klein and Nguyen \cite{KN13b}, so we can simply quote their result.

In fact, both Theorem
\ref{thm:W1} and Theorem \ref{thm:W2} can be proved in similar manner, with the help of the
condition \RCM, but this would result in less optimal volume dependence in Theorem \ref{thm:W1}.
As to Theorem \ref{thm:W2}, there is at present no basis of comparison -- as far as
arbitrary pairs of \emph{norm}-distant (and not Hausdorff-distant) cubes are concerned,
although in the author's opinion, the strange-looking exponent $2/3$ in the RHS of \eqref{eq:thm.W2}
is a mere artefact of the proposed method of proof.

\btm[Cf. \cite{KN13b}*{Theorem 2.2}]\label{thm:W1}
Let $\Sigma^{I^*}_{\Bx,L}$ be the spectrum $\Sigma(\BH_{\bcube_L(\Bx)})\cap I^*$, $I^* = [0,E^*]$. Then
\be\label{eq:thm.W1}
\pr{ \dist\big[ \Sigma^{I^*}_{\Bx,L}  , E\big] \le s } \le C_1(N,E^*, p_V) L^{Nd} \, s.
\ee
\etm

The estimate \eqref{eq:thm.W1} suffices for the fixed-energy analysis, but the derivation of dynamical and spectral localization
requires an EVC bound for pairs of local Hamiltonians (two-volume bound).

\btm\label{thm:W2}
Under the assumptions \Vone and \Uone, for any fixed $N,d$, $E^*$ and the PDF $F_V$
of the random scatterers,
there exist some $C_2, A_2\in(0,+\infty)$ such that for any pair of
$4NL$-distant cubes $\bcube_{L}(\Bx)$, $\bcube_{L}(\By)$ the following bound holds:
\be\label{eq:thm.W2}
\forall\, s\in(0,1] \quad
\pr{ \dist\big[ \Sigma^{I^*}_{\Bx,L}, \Sigma^{I^*}_{\By,L} \big] \le s } \le C_2 L^{A_2} \, s^{2/3}.
\ee
\etm

It is this EVC estimate which makes possible the present work.
Its proof relies on the following elements.

\begin{definition}
A  cube  $\bcube^{(N)}_L(\Bx)\subset (\DR^d)^N$ is weakly separated (or weakly $Q$-separated)
from $\bcube^{(N)}_L(\By)$ if
there exists a bounded subset $Q\subset \DR^d$,
of diameter $R \le 2NL$,  and the index subsets $\cJ_1, \cJ_2\subset [[1,N]]$  such that
$|\cJ_1| > |\cJ_2|$ (possibly, with $\cJ_2=\varnothing$) and
\begin{equation}\label{eq:cond.WS}
\begin{array}{l}
\big(  \Pi_{\cJ_1} \bcube_L(\Bx) \cup \Pi_{\cJ_2} \bcube_L(\By) \big) \;  \subseteq Q,\\
\big( \Pi_{\cJ^c_2} \bcube_L(\By) \cup \Pi_{\cJ^c_2} \bcube_L(\By) \big) \cap Q = \varnothing.
\end{array}
\end{equation}
A  pair of cubes $(\bcube_L(\Bx), \bcube_L(\By))$ is weakly separated if at least one of the cubes
is weakly separated from the other.
\end{definition}

In physical terms, the weak $Q$-separation of $\bcube_L(\Bx)$ from $\bcube_L(\By)$ means that there are more particles in $Q$
from the configurations $\Bu\in\bcube_L(\Bx)$ than from the configurations $\Bv\in\bcube_L(\By)$. This renders
the EVs of $\BH_{\bcube_L(\Bx)}$ more sensitive to the fluctuations of the random potential in $Q$
than the EVs of $\BH_{\bcube_L(\By)}$. Yet, one can still have $\rd^{(N)}_H(\bcube_L(\Bx), \bcube_L(\By))=0$,
which makes impossible any \emph{stricto sensu} stochastic decoupling of the random operators $\BH_{\bcube_L(\Bx)}(\om)$
and $\BH_{\bcube_L(\By)}(\om)$. This explains the choice of the term "weak" [separation].

\ble
Any pair of $N$-particle cubes $\bcube^{(N)}_L(\Bx)$, $\bcube^{(N)}_L(\By)$ with $\rd_S(\Bx,\By) > 4NL$
is weakly separated.
\ele

The proof is quite elementary and can be found in \cite{C10}.

Now the assertion of Theorem \ref{thm:W2} follows directly from the following result.

\ble\label{lem:WS.W2}

Let $V: \cZ\times \Omega \to \DR$ be a random field  satisfying the condition
\RCM. Consider the two weakly separated balls $\bball_{L}(\Bx)$, $\bball_{L}(\By)$ and the operators
$\BH_{\bball_{L}(\Bx)}(\omega)$, $\BH_{\bball_{L}(\By)}(\omega)$.
Then for any $s>0$ the following bound holds for their  spectra
$\Sigma_{\Bx,L}$, $\Sigma_{\By,L}$:
$$
\bal
\pr{ \dist(\Sigma_\Bx,\Sigma_(\By)) \le s }
\le |\bball_{L}(\Bx)| \, |\bball_{L}(\By)|\, C'L^{A'} (2s)^{b'} + C''L^{A''} (2s)^{b''}.
\eal
$$
\ele

\proof
Let $Q$ be a set satisfying the conditions \eqref{eq:cond.WS}
for some $\cJ_1, \cJ_2 \subset [[1,N]]$ with  $|\cJ_1| =: n_1 > n_2 := |\cJ_2|$.
Consider the sample mean $\xi=\xi_{Q}$ of $V$ over $Q$ and the fluctuations
$\{\eta_x, \, x\in Q \}$.
Owing to the flat
tiling\footnote{This is the only instance where the flat tiling is crucial for the two-volume EVC estimate.
The assumption $\rC_\ffi=1$ made in Sect.~\ref{ssec:cond.on.V} allows us to avoid this extra factor in the rest of the proof.}
condition on the form of the scatterers,
the operators $\BH_{\bball_{L'}(\Bx)}(\omega)$,  $\BH_{\bball_{L''}(\By)}(\omega)$ can be represented as follows:
\begin{equation}\label{eq:Ham.decomp}
\begin{array}{l}
\BH_{\bball_{L'}(\Bx)}(\omega) = n_1 \xi(\omega) \, \one + \BA(\omega), \\
\BH_{\bball_{L''}(\By)}(\omega) = n_2 \xi(\omega) \,\one + \BB(\omega),
\end{array}
\end{equation}
where the operators $\BA(\omega)$ and $\BB(\omega)$ are $\fF_{Q}$-measurable.
Spe\-cifically, let $\cJ_1^\rc = [[1,N]]\setminus\cJ_1$, $\cJ_2^\rc = [[1,N]]\setminus\cJ_2$,
and
$$
\bal
\BA(\omega) &= \BDelta + \BU_{\bball_{L'}(\Bx)} + \sum_{j\in \cJ_1^\rc} V(x_j;\om)
+ \sum_{j\in \cJ_1} \eta_{x_j}(\om),
\\
\BB(\omega) &= \BDelta + \BU_{\bball_{L''}(\By)} + \sum_{j\in \cJ_2^\rc} V(y_j;\om)
+ \sum_{j\in \cJ_2} \eta_{y_j}(\om) .
\eal
$$
Now \eqref{eq:Ham.decomp} follows from the identities
$$
\bal
V(x_j;\om) &= \xi(\om) + \eta_{x_j}(\om), \;\; j\in\cJ_1,
\\
V(y_j;\om) &= \xi(\om) + \eta_{y_j}(\om), \;\; j\in\cJ_2,
\eal
$$
since $\Pi_{\cJ_1}\bball_{L'}(\Bx), \Pi_{\cJ_2}\bball_{L''}(\By) \subset Q$,
$|\cJ_1|=n_1$, $|\cJ_2|=n_2$.

Let
$\{ \lambda_1, \ldots, \lambda_{M'}\}$ and
$\{ \mu_1, \ldots, \mu_{M''}\}$, with $M' = \,\,|\bball_{L'}(\Bx)|$,
$M'' = |\bball_{L''}(\By)|$,
be the sets of eigenvalues of $\BH_{\bball_{L'}(\Bx)}$ and of $\BH_{\bball_{L''}(\By)})$,
counting multiplicities.
By \eqref{eq:Ham.decomp}, these eigenvalues can be represented as follows:
$$
\bal
\lambda_j(\omega) = n_1\xi(\omega) + \lambda_j^{(0)}(\omega), \;\;
%
\mu_j(\omega) = n_2\xi(\omega) + \mu_j^{(0)}(\omega),
\eal
$$
where the random variables
$\lambda_j^{(0)}(\omega)$ and $\mu_j^{(0)}(\omega)$ are $\fF_{Q}$-measurable. Therefore,
$$
\lambda_i(\omega) - \mu_j(\omega) =  (n_1-n_2)\xi(\omega) + (\lambda_j^{(0)}(\omega) -  \mu_j^{(0)}(\omega)),
$$
with $n_1-n_2 \ge 1$, by our assumption.
Further, we can write
$$
\bal
\pr{ \dist(\Sigma_\Bx, \Sigma_\By) \le s }
& \le \sum_{i=1}^{M'} \sum_{j=1}^{M''}
     \bigesm{ \pr{ |\lambda_i - \mu_j| \le s \,| \fF_{Q}}}.
\eal
$$
Note that for all $i$ and $j$ we have
$$
\bal
\pr{ |\lambda_i - \mu_j| \le s \,|\, \fF_{Q}}
& = \pr{ |(n_1 - n_2)\xi + \lambda_i^{(0)} - \mu_j^{(0)}| \le s \,| \fF_{Q}}
\\
& \le \sup_{t\in\DR}  \pr{ |\xi -t| \le (n_1 - n_2)^{-1}s \,| \fF_{Q}}
\\
&\le  \sup_{t\in\DR} \big(  F_\xi(t + s \,| \fF_{Q}) -  F_\xi(t -  s \,| \fF_{Q})  \big)
\eal
$$
(we used $(n_1 - n_2)^{-1} \le 1$).
Consider the event
$$
\cE_L = \Bigl\{\; \sup_{t\in\DR} \;
\big|F_\xi(t+2s\,| \fF_{Q}) - F_\xi(t\,| \fF_{Q})\big|\ge C'L^{A'} (2s)^{b'} \Bigr\}.
$$
By \RCM, $\pr{\cE_L} \le C''L^{A''} (2s)^{b''}\}$.
Therefore,
$$
\bal
&{\pr{ \dist(\Sigma_\Bx, \Sigma_\By)) \le s }
= \esm{ \pr{  \dist(\Sigma_\Bx, \Sigma_\By) \le s  \,| \fF_{Q}}}}
\\
&\qquad {\le \esm{ \one_{\cE^c_L} \pr{ \dist(\Sigma_\Bx, \Sigma_\By) \le s \,| \fF_{Q}}}
+ \pr{\cE_L}}
\\
& \qquad{\le |\bball_{L''}(\Bx)| \cdot |\bball_{L''}(\By)|\, C'L^{A'} (2s)^{b'}
+ C''L^{A''} (2s)^{b''}.}
\eal
$$
\qedhere

The two-volume EVC estimate \eqref{eq:thm.W2} is thus established.

It is readily seen that a more traditional, one-volume EVC bound can be proved with
an analogous (indeed,  simpler) argument.

\section{Fixed-energy analysis of the flat tiling alloy model} \label{sec:FEMPMSA}

\subsection{The almost sure spectrum}

The exact location of the a.s. spectrum of the $N$-particle Hamiltonian $\BH^N(\om)$ in the entire Euclidean space $\DR^d$
can be easily found with the help of the classical Weyl criterium. The flat tiling alloy
is a particular case of a more general one studied by Klein and Nguyen,
and the only point which prevents us from  quoting  their result (cf. \cite{KN13b}*{Proposition A.1})
is that they considered an interaction of finite range. A careful reading of the proof given in \cite{KN13b}
evidences that this is a pure formality, for their argument, based on the Weyl criterium, naturally extends to any interaction potential
decaying at infinity. With these considerations in mind, we come to the following characterization
of the a.s. spectrum.

\bpr[]\label{prop:spec.R+}
Under the assumptions \Vone and \Uone, with probability $1$, $\Sigma(\BH^N_{\bcX^N}(\om)) = [0,+\infty)$.
\epr

With $N^*\ge 2$ fixed, we will prove, as in \cite{KN13b}, localization bounds in an energy interval $I^*=[0, E^*]$
with $E^*>0$ determined by the parameters of the model. Specifically, for any given (common) marginal PDF $F_V$ of the scatterers'
amplitudes satisfying \Vone, we can guarantee that our bounds, implying exponential spectral and sub-exponential dynamical
localization, hold true in an interval $I^*$ of positive length; the starting point for the scaling analysis is, as usual,
a Lifshitz tail estimate. Furthermore, for the potential $gV(x;\om)$ with fixed PDF $F_V$ and $g>0$ large enough, the large deviations
estimate can be replaced with a much simpler probabilistic argument,
gong back to \cite{DK89} and proving the ILS bound for \emph{any continuous} $F_V$ and $g\gg 1$.

As was said, we consider the finite-volume analysis more relevant for applications to physical models; keeping this in mind,
note that the spectrum of $\BH^{(N)}_{\Lam^N}(\om)$ is of course random, with the ground state energy $E^{(N)}_0(\Lam)$
positive with probability $1$, but, clearly, $E^{(N)}_0(\Lam) \to 0$ in probability, as $\Lam\nearrow \DR^d$. Therefore,
localization bounds established even in a tiny interval $[0,E^*]$, $0 < E^* \ll 1$, make sense for all $\Lam$ large enough.

\subsection{Dominated decay of the GFs}

The main technical tool used here is the Geometric Resolvent Inequality, well-known in the single-particle theory and
applicable to the multi-particle Anderson Hamiltonians as well, for the structure of the potential is irrelevant for this
general analytic result.

\bpr[Cf. \cite{St01}*{Lemma 2.5.4}]
Let be given two cubes $\bcube = \bcube_\ell(\Bu) \Subset \bcube' = \bcube_L(\Bv)$.
There is a real number $\Cgeom$ depending upon $\dist(\bcube, \pt \bcube')$, such that
for any measurable sets
$\ibA\subset \bcube_{L/3}$ and $\ibB \subset \bcube' \setminus\bcube$,
\be\label{eq:GRI.1}
\| \one_{\ibB} \BG_{\bcube}(E) \one_{\ibA} \|
\le \Cgeom
   \| \one_{\ibB} \BG_{\bcube'}(E) \one_{\bcubeout} \| \cdot \| \one_{\bcubeout} \BG_{\bcube}(E) \one_{\ibA} \|
\ee
\epr

Here and below, the superscript "out" refers to the (internal) $1$-neighborhood of the boundary of a given cube.

Introduce a notation that will be often used below:
\be
\dnorm{ \BG_{\bcube_L(\Bu)} } := \big\| \one_{\bcubeout_L(\Bu)} \BG_{\bcube_L(\Bu)} \chi_\Bu \big\| ,
\ee
(here $\curlywedge$ symbolizes the decay from the center to the boundary of a cube),

We define the external boundary $\pt \bball_L(\Bu)  = \{ \By\in\bball_L(\Bu):\, |\Bu - \By|= L\}$
in such a way that
\be\label{eq:def.pt.bball}
\bcube_{L}(\Bu) \subset \bigcup_{\Bx\in\pt \bball_L(\Bu)} \BC(\Bx).
\ee

\bco
Consider the embedded cubes $\bcube=\bcube_{L_k}(\Bx) \subset \tbcube=\bcubkoneu$ with $L_{k+1} .. L_k$.
Let $\tbball = \tbcube \cap \bcZ$,
For any cell $\BC(\By)\subset\bcube_{L_k}^{(out)}$, one has
\be
\| \chi_\By \BG_{\tbcube} \chi_\Bu \| \le \Cgeom \, \dnorm{\BG_{\bcube}} \; \| \chi_\By \BG_{\tbcube} \one_{\bcube^{out}} \|,
\ee
and consequently (cf. \eqref{eq:def.pt.bball}),
\be
\bal
\| \chi_\By \BG_{\tbcube} \chi_\Bu \| &\le \sum_{\Bz\in \pt \bball} \Cgeom \, \dnorm{\BG_{\bcube}} \; \| \chi_\By \BG_{\tbcube} \chi_\Bz \|
\\
& \le
  C' \, L^{Nd} \dnorm{\BG_{\bcube}} \; \max_{\Bz\in \pt \bball} \| \chi_\By \BG_{\tbcube} \chi_{\Bz} \|
\eal
\ee
with $C' = C'(N,d,\Cgeom)$.
\eco

Introduce the function $f_\By: \tbball \mapsto \DR_+$ by
\be\label{eq:cond.domin.2}
f_\By: \, \Bx \mapsto \| \chi_\By \BG_{\tbcube} \chi_{\Bx} \|,
\ee
and assume that
\be\label{eq:cond.domin.1}
C' \, L^{Nd} \dnorm{\BG_{\bcube_{L_k}(\Bx)}} \le q < 1,
\ee
then
\be
f_\By(\Bx) \le q \, \max_{\Bz\in \pt \bball} f_{\By}(\Bz).
\ee
Observe that $|\Bz - \Bx|=L_k$ for all $\Bz\in\pt \bball$. In the case where \eqref{eq:cond.domin.1} is fulfilled
for all cubes $\bcube_{L_k}(\Bx) \Subset \bcube_{L_{k+1}}(\Bu)$, one can iterate the above inequality and obtain
the following estimate (see the details in Appendix \ref{app:dominated}):
\be\label{eq:bound.domin.no.NS.1}
f_\By(\Bv) \le q^{\left\lfloor\frac{L_{k+1}}{L_k}\right\rfloor} \max_{\Bx \in\bball_{L_{k+1}}(\Bv)} f_\By(\Bx).
\ee

Let $\Sigma_{\Bv,L_{k+1}}$ be the spectrum (as a set) of the operator $\BH_{\bcube_{L_{k+1}}}(\Bv)$.
Then it is straightforward that, for $E\not\in \Sigma_{\Bv,L_{k+1}}$,
\be
 \max_{\Bx \in\bball_{L_{k+1}}(\Bv)} f_\By(\Bx) \le \| \BG_{\bcube_{L_{k+1}}(\Bv)}(E)\|
 \le \left( \dist\big( \Sigma_{\Bv,L_{k+1}}, \, E \big)\right)^{-1}.
\ee

In some instances of the scaling analysis, the single-step application of the GRI \eqref{eq:GRI.1}
would not provide a required (exponential or sub-exponential) decay of the GFs, so one has to make a few iterations.

Suppose we are in a situation where a given cube $\bcube_{L_k}(\Bx)$ does not satisfy the condition \eqref{eq:cond.domin.1},
so the value of the function $f_\By$ at the point $\Bx$ is not dominated by the maximum of its values
over the spheric layer of radius $L_k$, centered at $\Bx$. There is however a possibility to bound $f_\By(\Bx)$ in a slightly different way.
Consider a cube $\bcube_{L}(\Bv)\supset \bcube_{L_k}(\Bx)$ and assume that $E$ is not an eigenvalue of $\BH_{\bcube_{L}(\Bv)}$.
Then
\be
\bal
\| \chi_\By \BG_{\tbcube} \chi_\Bx \| &
\le \| \chi_\By \BG_{\tbcube} \one_{\bcube^{out}_{L}(\Bv)}\| \; \| \one_{\bcube^{out}_{L}(\Bv)} \BG_{\bcube_{L}(\Bv)} \one_\Bx\|
\\
& \le C L_k^{Nd}  \vertii{  \BG_{\bcubeLk} }
\; \max_{\Bu\in \bcube^{out}_{L}(\Bv)\cap \bcZ }
\| \chi_\By \BG_{\bcubeLkone} \chi_\Bu \| .
\eal
\ee
Further, assume in addition that for all $\Bu\in \bcube^{out}_{L}(\Bv)\cap \bcZ$ we do have the bound of the form
\eqref{eq:cond.domin.2} (with $\Bx$ replaced with $\Bu$), then we obtain
\be
\bal
f_\By(\Bx) &\le | \bcube^{out}_{L}(\Bv)\cap \bcZ| \cdot q \max_{\Bz\in \bcube_{L+L_k}(\Bv)\cap \bcZ } f_\By(\Bz)
\\
& \le q' \max_{\Bz\in \bcube_{L+L_k}(\Bv)\cap \bcZ } f_\By(\Bz).
\eal
\ee
with $q' = C L^{Nd} q$; in the course of the scaling analysis, $q$ (hence $q'$, too) will be of the form $\eu^{-aL_k^\delta}$, $a,\delta>0$,
which makes the polynomial factors $L^{Nd}$ fairly harmless.

Combining the above two procedures,
we will prove in Appendix \ref{app:dominated} an analog of the bound \eqref{eq:bound.domin.no.NS.1}.

The above discussion leads us to the following
\bde
Let be given real numbers $m>0$, $\delta\in(0,1]$, $E$ and integers $k>0$, $N\ge 1$.
A cube $\bcubeN_{L_k}(\Bu)$, as well as its lattice counterpart $\bballN_{L_k}(\Bu)$, is called
$(E,\delta,m)$-non-singular ($(E,\delta,m)$-NS) if
\be
 \Cgeom\, (3L_k)^{Nd} \, \dnorm{ \BG_{\bcube_{L_k}\Bx}(E)} \le \eu^{ - m L_k^{\delta}}
\ee

$\bcubeN_{L_{k+1}}(\Bu)$, as well $\bballN_{L_{k+1}}(\Bu)$, is called
$(E,\beta)$-NR if
\be
 \dist\big( \Sigma_{\Bv,L}, \, E \big) \ge \eu^{-L^\beta}
\ee
$\bcubeN_L(\Bu)$ and $\bballN_L(\Bu)$ are called E-CNR if for all $L_k \le \ell \le L_{k+1} - L_k$
\be
 \dist\big( \Sigma_{\Bv,L}, \, E \big) \ge \eu^{-L_{k+1}^\beta} .
\ee
\ede

Here $(3L)^{Nd}$ is a (crude) upper bound on the cardinality $| \pt^- \bball_L(\Bx)|$.

\subsection{Induction hypothesis}
The goal of the scale induction is to prove recursively the following property:

\SS{N,k}:
Given integers $N^*\ge 3$, $L_0\ge 1$, $Y\ge 2$, the integer sequence $\{L_j := L_0 Y^j,\; j\ge 0\}$,
the real numbers $m^*>0$, $\nu^*>0$ and the finite sequences
$$
m_n := m^*(1 + 4L_0^{-\delta+\beta})^{N^*-n}, \;\; \nu_n := \nu^*(2Y^\kappa)^{N^*-n}, \;
1 \le n \le N^*,
$$
the following property is fulfilled for all $1 \le n \le N$:
\be
\pr{ \bcube^{(n)}_{L_k}(\Bx) \text{ is $(E,\delta,m_n)$-S } } \le \eu^{ - \nu_n L_k^\kappa}.
\ee

\subsection{Initial length scale estimate}

The assumption of non-negativity of the interaction greatly simplifies the EVC analysis in the continuous multi-particle models
near the bottom of the spectrum.
The key observation here is that any non-negative interaction can only move the EVs up, thus resulting automatically in stronger
ILS estimates (in any interval of the form $(-\infty, E^*]$)
for the interactive model at hand than with the interaction switched off.

We cannot apply directly the ILS estimate from \cite{CBS11}*{Lemma 3.1}, for the latter provides only a power-law
decay of the probability of unwanted events, while we need an input for the sub-exponential MSA
induction\footnote{This is the price to pay for skipping the first phase of the bootstrap in our simplified scheme.}.
However,
a direct inspection of the proof of Theorem 2.2.3 in \cite{St01}, on which Lemma 3.1 in  \cite{CBS11} is based,
shows that for any $0 < \gamma < 1/2$ and $\eps>0$ there exist $L_0 >0$, $m, \nu\in,(0,+\infty)$ such that
$$
\bal
\pr{ \|\one_{\bcube^{(\rm out)}_L(\Bx)}  \BG_{\bcube_L(\Bx)}(E) \chi_x \| > \eu^{ - m L^{\frac{1+\gamma}{2}}} }
\le \eu^{-\nu L_k^{\frac{1-\gamma -\eps}{2}}}
\eal
$$
or, equivalently,
$$
\bal
\pr{ \|\one_{\bcube^{(\rm out)}_L(\Bx)}  \BG_{\cube_L(x)}(E) \chi_\Bx \| > \eu^{ - (m L^{\eps/2}) L^{\frac{1+\gamma-\eps}{2}}} }
\le \eu^{-(\nu L^{\eps/2}) L^{\frac{1-\gamma -2\eps}{2}}}
\eal
$$
For example, with $\gamma = 1/12$  we obtain
$$
\bal
\pr{ \|\one_{\bcube^{(\rm out)}_{L_0}(\Bx)}  \BG^{(1)}_{\bcube_{L_0}(\Bx)}(E) \chi_\Bx \| > \eu^{ - m(L_0) L^{2/3}} }  \le \eu^{- \nu(L_0) L_0^{1/4}} .
\eal
$$
where $m_0(L_0)$, $\nu_0(L_0)\to+\infty$ as $L_0\to+\infty$.

The bootstrap strategy ultimately results in stronger estimates following from weaker initial assumptions,
but extracting such estimates requires one to go through the bootstrap steps, which takes a bit longer
than a more straightforward approach summarized, e.g., in the book \cite{St01}.

Summarising, we come to the following
\bpr
For some $\kappa>0$ and any $m^*, \nu^*\ge 1$ there exists an integer $L^*_0 = L^*_0(m^*, \nu^*, N^*)$ such that
\SS{N,0} holds true for all $1\le N \le N^*$ and $L_0\ge L^*_0$.
\epr

\subsection{Analytic scaling step}

\bde
A cube $\bcubeN_{L_k}(\Bu)$ is called weakly interactive (WI) if
$$
\diam \Pi \Bu \equiv \max_{i\ne j} |u_i - u_j | \ge 3NL_k,
$$
and strongly interactive (SI), otherwise.
\ede

Observe that the properties WI/SI are permutation-invariant, so that both the norm-distance and its symmetrized counterpart
$\rd_S$ can be used in the next definition.

\bde
A cube
$\bcubeN_{L_{k+1}}(\Bx)$ is called \Ebad if it contains either a weakly interactive \EdS cube of radius $L_k$ or a pair of $9NL_k$-distant,
\EdS, strongly interactive cubes of radius $L_k$. Otherwise, it is called \Egood.
\ede

For the reader's convenience, we summarize in the table below the assumptions on the key parameters used
in the scale induction with $L_k = L_0 Y^k$, $k\ge 0$.

\def\tabup{\rule{0pt}{3ex}}
\def\tabdown{\rule[-2ex]{0pt}{0pt}}

\renewcommand{\arraystretch}{1.2}
\be\label{eq:table1}\hbox{\begin{tabular}{|c|c|}
  \hline
\tabup
   $0 < \kappa <\beta < \delta < \min\big[ \zeta, 1 \big]  $\,
      &  $Y \ge \max\left[  24 N^*, \, 12^{\frac{1}{1 - \delta}} \right], \; \text{so }\,  \quart Y^{1-\delta} \ge  3  $
\tabdown
 \\
  \hline
\tabup
    $m_N = m^*\,\big(1 + 4L_0^{-\delta+\beta}\big)^{N^*-N}$ &
       $\begin{array}{ll}\nu_N = \nu^*\, (2 Y^\kappa)^{N^*-N}\\ \end{array}$
\tabdown
\\
\hline
\end{tabular}}\ee

The pivot of the deterministic component of the scaling analysis
is the following result, well-known in the single-particle theory.

\ble\label{lem:good.NR.is.NS}
Fix the integer $Y>1$ and suppose that a cube $\bcubeLkone(\Bx)$ is \Egood and {\rm\ENR}.
If $L_0$ is large enough, then $\bcubeLkone(\Bx)$ is {\rm\EdNS}.
\ele

For completeness, we sketch the proof in Appendix \ref{app:dominated}.

\subsection{Probabilistic scaling step}

\subsubsection{Weakly interactive cubes }

\ble\label{lem:WI.decomp}
For any weakly interactive cube $\bcubeN_{L_k}(\Bu)$ there is a decomposition
$\bcubeN_{L_k}(\Bu) = \bcube^{n'}_{L_k}(\Bu') \times \bcube^{n''}_{L_k}(\Bu'')$
with
\be
\dist\big( \Pi \bcube^{n'}_{L_k}(\Bu'), \, \Pi \bcube^{n''}_{L_k}(\Bu'') \big) > L.
\ee
\ele

\proof
Assuming $\diam( \Pi \Bu) > 3 N L$, let us show that the projection $\Pi \bcube(\Bu,3L/2)$ is a disconnected subset of $\cZ$.
Assume otherwise; then for any partition $\cJ \sqcup \cJ^\rc = \{1, \ldots, N\}$,
we have $\rd(\Pi_\cJ \Bu, \Pi_{\cJ^\rc}\Bu)$   $\le 2 \cdot \frac{3L}{2}=3L$. An
induction in $N\ge 2$ shows that
$\diam\; \Pi\Bu \le (N-1) \cdot 3L < 3NL$, contrary to our hypothesis.

Therefore, we have
$\rd\left( \Pi_\cJ \bcube_{3L/2}(\Bu),  \Pi_{\cJ^\rc} \bcube_{3L/2}(\Bu)\right) > 0$,
for some partition $(\cJ,\cJ^\rc)$,
hence
$$
\rd\left( \Pi_\cJ \bcube_{3L/2}(\Bu),  \Pi_{\cJ^\rc} \bcube_{3L/2}(\Bu)\right)> {\textstyle \frac{1}{2}} L + {\textstyle\frac{1}{2}} L = L ,
$$
as asserted.
\qedhere

We will assume that one such decomposition is associated with each WI cube (even if it is not unique), and call
it the canonical one. For the Hamiltonian in a WI cube we have the following algebraic representation:
with $\bcube' = \bcube^{n'}_{L_k}(\Bu')$, $\bcube''=\bcube^{n''}_{L_k}(\Bu'')$,
\be
\bal
\BH &= \BHni + \BU_{\bcube', \bcube''}
\\
& = \BH_{\bcube'} \otimes \one^{(n'')} + \one^{(n')} \otimes \BH_{\bcube''} + \BU_{\bcube', \bcube''}
\eal
\ee
where, due to the assumption \Uone,
\be\label{eq:Uone.norm.Ubcube.bcube}
\| \BU_{\bcube', \bcube''} \| \le C \eu^{-L_k^\zeta}.
\ee

\ble\label{lem:prob.WI.S}
Assume the property {\rm\SS{N-1,k}}. If $L_0$ is large enough, then for any WI cube $\bcubeN_{L_k}(\Bu)$
\be\label{eq:lem.prob.WI.S.1}
\pr{ \bcubeN_{L_k}(\Bu) \text{ is $(E,\delta, m_N)$-S}} \le \eu^{ - \frac{3}{2} \nu_N L_{k+1}^{\kappa}}
\ee
and therefore,
\be\label{eq:lem.prob.WI.S.2}
\bal
\pr{ \bcubeN_{L_k}(\Bu) \text{ contains  a WI $(E,\delta, m_N)$-S ball of radius $L_k$}}
\le \frac{1}{4} \eu^{ - \nu_N L_{k+1}^{\kappa}} .
\eal
\ee
\ele

See the proof in Appendix \ref{app:proof.WI.S}.

\subsubsection{Strongly interactive cubes}

\ble\label{lem:8NL.dist}
If two {\rm SI} cubes $\bcubeN_L(\Bx)$, $\bcubeN_L(\By)$ are $9NL$-distant
and $L > 2\rr_0$, then
\be
\Pi \bcubeN_{L+\rr_0}(\Bx) \cap \Pi \bcubeN_{L+\rr_0}(\By) = \vempty
\ee
and, consequently, the random operators $\BH_{\bcubeN_L(\Bx)}$ and $\BH_{\bcubeN_L(\By)}$
are independent.
\ele

\proof
By definition, for any SI cubes $\bcube^{(N)}_{L}(\Bx)$, $\bcube^{(N)}_{L}(\By)$
we have
$$
\max_{i,j} \rd(x_i, x_j) \le 3NL, \;\; \max_{i,j} \rd(y_i, y_j) \le 3NL,
$$
and it follows from the assumption $\rd(\Bx,\By) > 9 NL$ that
for some $i',j'\in\{1, \ldots, N\}$
$
\rd(x_{i'}, y_{j'}) > 3NL,
$
thus for any $i,j\in\{1, \ldots, N\}$
$$
\bal
\rd(x_i, y_j) \ge \rd(x_{i'}, y_{j'}) - \rd(x_{i'}, x_{i'}) - \rd(x_{j'}, y_{j})
> 9NL - 6NL - 2\rr_0 \ge 2NL.
\eal
$$
Therefore,
$$
\dist\big(\Pi \bcube_{L+\rr_0}(\Bx), \Pi \bcube_{L+\rr_0}(\By) \big) > 2(N-1) L \ge 0,
$$
so $\Pi \bcube^{(N)}_{L+\rr_0}(\Bx) \cap \Pi \bcube^{(N)}_{L+\rr_0}(\By) = \varnothing$.
This implies independence of the samples of the random potential in   $\BH_{\bcube^{(N)}_{L+\rr_0}(\Bx)}(\om)$
and $\BH_{\bcube^{(N)}_{L+\rr_0}(\By)}(\om)$.
\qedhere

\subsubsection{The scale induction}

\btm
Suppose that \SS{N,0} holds true, and for all $k\ge 0$, one has
$$
\pr{ \text{ $\bcube_{L_{k+1}}(\Bu)$ is \ER } } \le \quart \eu^{ - \nu_N L_k^\kappa},
$$
for some $\kappa < \beta<\delta$.
If $L_0$ is large enough, then \SS{N,k} holds true for all $k\ge 0$.
\etm

\proof
It suffices to derive \SS{N,k+1} from \SS{N,k}.
By Lemma \ref{lem:good.NR.is.NS}, if $\bcube_{L_{k+1}}(\Bu)$ is $(E,\delta,M)$-S, then it is either $(E,\beta)$-R
or \Ebad. Let
\begin{align}
\notag
\rP_i &:= \pr{ \text{ $\bcube_{L_{k+1}}(\Bu)$ is \EdS} }, \;\; i=k, k+1,
\\
\notag
\rS_{k+1} &:= \pr{ \text{ $\bcube_{L_{k+1}}(\Bu)$ contains a WI, \EdS cube of radius $L_k$} },
\\
\label{eq:Q.k+1}
\rQ_{k+1} &:= \pr{ \text{ $\bcube_{L_{k+1}}(\Bu)$ is \ER } } \le \quart \eu^{ - \nu_N L_k^\kappa},
\end{align}
(the last inequality is assumed, but its validity actually follows from Theorem \ref{thm:W1}).
Further, an \Ebad cube $\bcube_{L_{k+1}}(\Bu)$ must contain either a WI, \EdmS cube of radius
$L_k$ (with probability
$\rS_{k+1}\le \quart \eu^{-\nu_N L^\kappa_{k+1}}$
by Lemma \ref{lem:prob.WI.S}), or at least one pair of $9NL_k$-distant cubes
$\bcube_{L_k}(\Bv_i)$, $i=1,2$, which are \EdmS. By virtue of Lemma \ref{lem:8NL.dist}, the random operators
$\BH_{\bcube_{L_k}(\Bv_1)}(\om)$, $\BH_{\bcube_{L_k}(\Bv_2)}(\om)$ are independent, thus such a pair
inside $\bcube_{L_{k+1}}(\Bu)$ exists with probability
$$
\le C L_{k+1}^{2 Nd} \rP_k^{2}
\le \eu^{ - 2 \nu_N L_k^\kappa  + C\ln L_k}
 \le \quart \eu^{ - \nu_N L_k^\kappa} ,
$$
provided $L_0$ (hence every $L_k$, $k\ge 0$) is large enough.
Therefore,
$$
\bal
\rP_{k+1} &\le C L_{k+1}^{2 Nd} \rP_k^{2} + \rS_{k+1}+  Q_{k+1}
\\
& \le \quart \eu^{ - \nu_N L_k^\kappa} + \quart \eu^{ - \nu_N L_k^\kappa} + \quart \eu^{ - \nu_N L_k^\kappa}
< \eu^{ - \nu_N L_k^\kappa} .
\eal
$$

\qedhere

This marks the end of the fixed-energy MPMSA.

\section{Derivation of strong dynamical localization}
\label{sec:reduction}

\subsection{Variable-energy MPMSA estimates}
\label{ssec:VEMPMSA.exp}

We need an adaptation to the Schr\"{o}din\-ger operators in a Euclidean space
of a fairly general statement presented in \cite{C14a} which encapsulates a result by Elgart et al.
\cite{ETV10}. Pictorially, it says that with high probability, the (random) energy set dangerous for localization is thin.

It seems appropriate to remind here that the idea
of using the augmented disorder-energy measure space
$\Om\times I$ along with the Chebychev inequality and the Fubini theorem has been employed long ago
by Martinelli and Scoppola \cite{MS83} who derived from the results of the fixed-energy MSA the a.s. absence
of a.c. spectrum. Following Elgart et al. \cite{ETV10}, we use a similar strategy to infer -- in a ``soft'' way -- from the results of the fixed-energy
analysis much stronger properties: absence of s.c. spectrum (spectral localization) and strong dynamical
localizaiton, with sub-exponential decay of the EFCs.

\bpr\label{prop:ETV}
Let be given a cube $\bcube =\bcube^{(N)}(\Bu,L)$, $L>0$, and the random operator
$\B{H}_\bcube=\B{H}^{(N)}_\bcube(\om)$ of the form
\be\label{HamBSA}
\left(\BH_\bcube f\right)(\Bx )=\left(-\Delta_\bcube f\right)(\Bx)+W(\Bx;\om )f(\Bx),
\;\;\Bx\in\bcube ,\ee
where $(\Bx,\om )\mapsto W(\Bx;\om )\in\DR$ is a given random potential energy.
Denote
$$
\BF_\Bu(E) = \max_{\Bz\in \bcube^{(out)}_L(\Bu)\cap \bcZ} \langle \chi_\Bz \BG_{\bcube_L(\Bu)}(E) \chi_\Bu \rangle
$$

Consider  an interval $I\subset\DR$ and denote by $E_j=E_j(\om )$, $1\leq j\leq M$,
be the (random) eigenvalues of $\BH_{\bcube_L(\Bu)}$ in $I$ listed in increasing order.
Denote $\Sigma_{\Bu,L}^I = \Sigma(\BH_{\bcube_L(\Bu)})\cap I$.
Let the numbers $a=a_L, b=b_L, c=c_L, q_L>0$ satisfy
\begin{align}\label{eq:cond.aL.bL.cL}
b_L \le \min\{ a_L c_L^2, \, c_L\} ,
\\
\label{eq:cond.qL}
\sup_{E\in I} \, \pr{ \BF_\Bu(E) \ge a_L } \le q_L .
\end{align}
Then there is an event $\cB$ of probability
$\pr{\cB}\le |I| b_L^{-1} q_L$ such that
for all $\om\not\in\cB$
\be\label{eq:def.cE}
\csE_\Bu(2a_L) := \left\{ E\in I:\; \BF_\Bu(E) > 2a_L\right\} \subset \cup_{j=1}^{M} I_j,
\ee
where $I_j := (E_j-2c_L,E_j+2c_L)$.
\epr

\proof
Let $\csE(a) = \{E\in I:\, \BF_\Bu(E) \ge a\}$,
$\cB_\Bu(b) = \{\om:\, \mes\, \csE(a) \ge b\}$.
By Chebyshev's inequality and the Fubini theorem, it follows from the assumption \eqref{eq:cond.qL}
that
$$
\bal
\pr{ \cB_\Bu(b) } & \le b^{-1} \esm{ \mes\, \csA(a)} = b^{-1} \int_I dE\, \esm{ \one_{\{\BF_\Bu(E) \ge a\}}}
   \le b^{-1} \, |I| \, q_L.
\eal
$$
Fix any $\om\not\in\cB(b)$, so that $\mes(\csE(a)) < b$. Let
$\csR(r) = \{E\in I:\, \dist(E, \Sigma^I_{\Bu,L} \ge r\}$, for  $r\ge 0$. Observe that
$\csA(b) := \{E\in I:\, \dist(E, \csR(2c) <b\}\subset \csR(c)$, and $\csA^\rc(b)$
is a union of some intervals at distance $\ge c$ from $\Sigma^I_{\Bu,L}$.

Let us show by contraposition that
$$
\forall \, \om\not\in \cB(b) \quad   \{E\in I:\, \BF_\Bu(E) \ge 2a\} \cap \csR(2c) = \varnothing.
$$
Assume otherwise and pick any $\lam\in \{E\in I:\, \BF_\Bu(E) \ge 2a\} \cap \csR(2c)$.
Let $J := \{E\in I:\, |E - \lam| < b\} \subset \csA_b \subset\csR(2c)$. By the first resolvent identity,
for any $E\in J$
$$
\bal
\| \BG(E) \| &\ge \| \BG(\lam) \| - |E - \lam| \, \| \BG(\lam) \| \, \| \BG(E) \|
\\
& \ge 2a - b \cdot c^{-1} \cdot c^{-1} \ge a,
\eal
$$
since $E,\lam\in \csR(c)$ and we assumed $b \le ac^2$. Thus $J\subset \csE(a)$
and $\mes\, \csE(a) \ge |J| \ge b$, which contradicts the definition of $\cB(b)$.

We conclude that for any $\om\not\in\cB(b)$, the set $\{E\in I:\, \BF_\Bu(E) \ge 2a\}$
is covered by the intervals of length $4c$ centered at the EVs $E_j\in I$.
\qedhere

\bco\label{cor:ETV}
Under the assumptions \Vone and \Uone,
for any $\nu>0$ there exists an interval $I^*=[0, E^*]$, with $E^*>0$, such that
for any $k\ge 0$ and any pair of $4NL_k$-distant cubes the following bound holds:
\be
\pr{\exists\, E\in I^*:\,  \text{$\bcube_{L_k}(\Bx)$ and $\bcube_{L_k}(\By)$ are $(E,\delta,m_N)$-S} }
\le \eu^{- \nu L^\kappa} .
\ee
\eco

\proof
We can apply Proposition \ref{prop:ETV} with $M \le L^C$, $C<\infty$ by Weyl's
law\footnote{Here Weyl's law provides a deterministic upper bound n the number of EVs in $I^*$, owing to
non-negativity of the potential energy; it suffices to consider $-\BDelta_{\bcube_L(\Bx)}$ and $-\BDelta_{\bcube_L(\By)}$.},
and
\be\label{eq:choice.aL.bL.cL}
a_L = \eu^{- \frac{\nu_N}{3} L^\kappa}, \; b_L = \eu^{- \frac{2\nu_N}{3} L^\kappa}, \; c_L = \eu^{- \frac{\nu_N}{7} L^\kappa},
\; q_L = \eu^{- \nu_N L^\kappa}.
\ee
Here $\nu_N>0$ can be made arbitrarily large by choosing $L_0$ large enough.
Let $\cB_\Bx$ and $\cB_\By$ be the events introduced in Proposition \ref{prop:ETV} relative to the cubes $\bcube_L(\Bx)$
and $\bcube_L(\By)$, respectively, and set $\cB = \cB_\Bx \cup \cB_\By$. Further,
define the energy sets $\csE_\Bx(\cdot)$ and $\csE_\By(\cdot)$ as in the LHS equation \eqref{eq:def.cE}, with $\Bu=\Bx$ and
$\Bu=\By$, respectively. Next, introduce the event
$\cbS_{\Bx,\By} = \{\om:\, \csE_\Bx(2a_L)\cap \csE_\By(2a_L) \ne \varnothing\}$. Then
\be\label{eq:ETV.prob.csS}
\pr{ \cbS_{\Bx,\By} } \le \pr{ \cbS_{\Bx,\By}\setminus\cB} + \pr{ \cB}
\le \pr{ \cbS_{\Bx,\By}\setminus\cB} + 2 b_L^{-1} q_L |I^*|.
\ee
For any $\om\not\in \cB$, each of the energy subsets $\csE_\Bx(2a_L)$, $\csE_\By(2a_L)$ is covered by the
intervals of length $4c_L$
centered at the respective EVs of $\BH_{\bcube_L(\Bx)}$ or $\BH_{\bcube_L(\By)}$. By the EVC estimate,
applied to the spectra $\Sigma^{I^*}_{\Bx,L}$, $\Sigma^{I^*}_{\By,L}$ of these operators within the interval $I^*$,
combined with Weyl's law, we have
\be\label{eq:ETV.prob.csS.minus.cB}
\pr{ \cbS_{\Bx,\By}\setminus\cB} \le \pr{ \dist\left( \Sigma^{I^*}_{\Bx,L}, \,\Sigma^{I^*}_{\By,L} \right) \le 4c_L }
\le C' L^{A'} c_L^{\th},
\ee
with $A' = A'(A, N, d)$.
Collecting \eqref{eq:choice.aL.bL.cL}, \eqref{eq:ETV.prob.csS} and \eqref{eq:ETV.prob.csS.minus.cB}, the claim follows.
\qedhere

\subsection{Decay of the EF correlators}

\bpr\label{lem:5.1}
{\rm{(Cf. \cite{C14a}*{Theorem 7})}}
\label{prop:GK}
Given a positive integer $L$, assume that the following bound holds true for a
pair of disjoint balls $\bcube_L(\Bx), \bcube_L(\By)\subset\bcZ$ and some positive functions $u, h$:
\be\label{eq:cond.thm.GK}
\pr{ \exists\, E\in \DR:\,
\min\left[ \BF_\Bx(E), \BF_\By(E) \right] > u(L)}\le h(L).\ee
Then for any finite connected subset $\BLam\supset\bcube_L(\Bx) \cup \bcube_L(\By)$ one has

\be\label{eq:thm.MSA.to.DL.sup}
\esm{ \sup_{t\in\DR}\;  \big|\langle\one_\Bx  \,|\, P_I\big(\BH_\bcube \big) \eu^{-\ii t \BH_\bcube} \,|\,  \one_\By \rangle \big| }
\le 4 u(L) + h(L).
\ee
\epr

We omit the proof which repeats almost verbatim that of Theorem 7 in \cite{C14a}, with minor adaptations
to continuous Schr\"{o}dinger operators; the key point of the proof is the Bessel inequality
(applied in \cite{C14a} to finite-dimensional local Hamiltonians $\BH_\bcube$) valid for the operators with compact resolvent.
The main idea goes back to the work by Germinet and Klein \cite{GK01} who simplified more involved
techniques from \cite{GDB98} and \cite{DS01}.

For a bounded $\BLam$, Theorem \ref{thm:main} now follows from

\btm\label{thm:GF.decay.Holder}
Given $ \nu >0$, $\exists$ $g_*=g_*(\nu)\in (0,\infty )$ and $C_*=C_*(\nu)\in (0,\infty )$ such that, for $|g|\ge g_*(\nu)$,
and $1\leq N\leq N^*$, $\forall$ $\Bx,\By\in\bcZ$ and a finite
$\BLam\subset\bcZ$ with $\BLam\ni\Bx,\By$,
\be\label{eq:dynlocinset}
\Upsilon_{\Bx,\By} :=
\esm{ \sup_{t\in\DR}\;  \big|\langle\one_\Bx  \,|\, P_I\big(\BH_\bcube \big) \eu^{-\ii t \BH_\bcube} \,|\,  \one_\By \rangle \big| }
\le  C_* \eu^{- \nu (\rd_{\rS}(\Bx,\By))^{\kappa}}.
\ee
\etm
\proof
Without loss of generality, it suffices to prove the assertion for the pairs of points with
$\rd_{\rS}(\Bx,\By)> 3NL_0$. Indeed, the EFC correlator is always bounded by $1$, so
for pairs $\Bx,\By$ with $\rd_{\rS}(\Bx,\By)\le 3NL_0$ the required decay bound
can be absorbed in a sufficiently large constant $C_*$.

Thus, fix points $\Bx,\By\in\bcZ$ with $R: =\rd_{\rS}(\Bx,\By)> 3NL_0$. There exists
$k$ such that $R\in(3NL_k, 3NL_{k+1}]$. Arguing as above, it suffices to consider
a finite $\BLam\subset\bcZ$ such that
$
\bball^{(N)}_{L_k}(\Bx) \cup \bball^{(N)}_{L_k}(\By) \subset \BLam .
$

Since
$R \le 3NL_{k+1} = 3N Y L_k$, we have $L_k \ge R/(3NY)$.
By Corollary \ref{cor:ETV} combined with Proposition \ref{prop:GK},
for any $\nu'>0$ and $|g|$ large enough,
\be\label{eq:locdynset}
\Upsilon_{\Bx,\By}
\le 4 \eu^{-\frac{m_N}{(3NY)^\delta} R^{\delta}} + \eu^{-\frac{\nu_N }{11(3NY)^\kappa} R^{\kappa}} .
\ee
Given an arbitrary $\nu>0$, choose a sufficiently large $L_0$, so that
the initial scale estimate  \SS{N,0} is fulfilled with
$m_N\ge 3NY \nu$, $\nu_N \ge 33 NY \nu$. Then we obtain
$$
\Upsilon_{\Bx,\By}
\le 5 \eu^{-\nu R^{\kappa}} = 5 \eu^{-\nu ( \rd_S(\Bx,\By) )^\kappa}
$$
This completes the proof of Theorem \ref{thm:GF.decay.Holder}.
\qedhere

An extension of the uniform bounds on the EFC decay to the unbounded domains $\BLam\subset\bcX^N$
can be obtained along the path laid down earlier by Aizenman et al. (cf., e.g., \cite{Ai94,ASFH01,AENSS06}), with the help
of the Fatou lemma on convergent measures; this lemma is applied to the spectral measures defined by the EF correlators.

\appendix

\section{Proof of Lemma \ref{lem:good.NR.is.NS}}
\label{app:dominated}

In this section, we deal with abstract finite connected graphs $\cG$ endowed with the graph distance $\rd = \rd_\cG$;
recall that $\rd_\cG(x,y)$ is the length of the shortest path from $x$ to $y$ over the edges of the graph.
A ball of radius $L$ centered at $u\in\cG$ is denoted $\ball_L(u)$. Given a funciton $f:\cG\to\DR$
and a subset $A\subset \cG$, we denote $\rM(f, A)  := \max_{x\in A} f(x)$.

Introduce the following notions.

\bde (1) Let be given two integers $L\ge \ell\ge 1$, a finite connected graph $\cG\supset\ball_L(u)$,
and a non-negative function $f:\, \ball_{L}(u)\to\DR_+$.
A point $x\in\ball_{L-\ell}(u)$ is called $(\ell,q)$-regular for the function $f$ if
\be
f(x) \le q \, \rM(f, \ball_{\ell}(x) ) .
\ee

The set of all regular points for $f$ is denoted by $\csR_f(u)$.

\noindent
(2) A spherical layer $\cL_r(u) = \{ y: |y - u| = r\}$ is called regular if $\cL_r(u)\subset \csR_f$.

\noindent
(3) For $x\in\ball_{L-\ell}(u)$, set
$$
r(x) :=
\begin{cases}
\min \{ r\ge |u-x|: \, \cL_r\subset \csR_f  \}, & \text{ if  there is } \cL_r\subset \csR_f \text{ with } r\ge |u-x| ,
\\
+\infty, & \text{ otherwise},
\end{cases}
$$
and
$R_f(x) = r(x) + \ell$.

\noindent
(4) Given a set $\csS \subset \ball'$, the function $f$ is called $(\ell,q,\csS)$-dominated in $\ball_L(u)$ if
$\ball_{L}(u) \setminus \csS \subset \csR_f(u)$, and for any $x\in\ball_{L-\ell}(u)$ with $R_f(x)<+\infty$, one has
$$
f(x) \le q\, \rM\big(f, \ball_{R_f(x)}(u) \big).
$$
\ede

The key feature of the $(\ell,q)$-dominated functions on discrete sets (in a more general context, finite graphs; cf.
\cite{CS13}) is the following result.

\ble\label{lem:f.domin}
Let a function $f:\cG\to\DR_+$ be $(\ell,q,\csS)$-dominated in a ball $\ball=\ball_L(u)$, with $L\ge \ell\ge 1$.
Assume that the set $\csS$ is covered by a union $\bcA$ of concentric annuli $\ball_{b_j}(u)\setminus \ball_{a_j-1}(u)$
with $\rw(\bcA) := \sum_j (b_j - a_j+1) \le L - \ell$. Then
$$
f(u) \le q^{ \frac{L - \ell - \rw(\bcA)}{\ell}} \rM\big(f, \ball_{L+1}(u)\big).
$$
\ele

\proof
It follows from the hypothesis that
$$
\frac{L - \rw(\bcA) }{ \ell } \ge \left\lfloor \frac{L - \rw(\bcA) }{ \ell } \right\rfloor =: n+1, \; n\ge 0.
$$

Define recursively a finite sequence of integers $\{r_n > r_{n-1} > \cdots > r_0 \}$:
\be\label{eq:def.rj}
\bal
r_n &= \max \big[ r \le L - \ell:\; \cL_r \cap \csS = \varnothing \big]
\\
r_j &= \max \big[ r \le r_{j+1} - \ell:\; \cL_r \cap \csS = \varnothing \big], \;\; j=n-1, \ldots, 0.
\eal
\ee
It is convenient to introduce also, formally, $r_{n+1} = L$, although the regularity property does not apply to the points in $\cL_{r_{n+1}}=\cL_L$.
Note that one can indeed construct in \eqref{eq:def.rj}
$n+1$ integers $r_j \ge 0$, since $L - \rw(\bcA) - (n+1)\ell \ge 0$, and we have
$$
L - r_0 = \sum_{j=0}^n (r_{j+1} - r_j) \le (n+1)\ell + \rw(\bcA),
$$
so $r_0 \ge L - \rw(\bcA) - (n+1)\ell \ge 0$.

Introduce the non-decreasing non-negative funciton
$$
F:\, r \mapsto \rM(f, \ball_r(u)), \;\; r\in\{0, 1, \ldots, L\}.
$$
For all $0 \le j \le n$, $r_j +\ell \le L$ and $\cL_{r_j}$ is regular, so
$$
F(r_n) = \rM(f, \ball_{r_n}(u)) \le q \rm(f, \ball_{r_n+\ell}(u)) \le q \rm(f, \ball)
$$
and for all $0 \le j \le n-1$,
$$
F(r_j) = \rM(f, \ball_{r_j}(u)) \le q \rm(f, \ball_{r_j+\ell}(u)) \le q F(r_{j+1}).
$$
Now the backward induction in $j=n, \ldots, 0$ proves the claim:
$$
\bal
f(u) &\le \rM(f, \ball_{r_0}(u)) = F(r_0) \le q^{n+1} \rM(f,\ball)
\\
& \le q^{ \left\lfloor \frac{L - \rw(\bcA)}{\ell} \right\rfloor} \rM(f, \ball)
\le q^{ \frac{L - \rw(\bcA) - \ell }{\ell} } \rM(f, \ball).
\eal
$$
\qedhere

The relevance of the notion of dominated decay is explained by the next result following immediately from the GRI.
\ble\label{lem:GF.is.domin}
Suppose that for some integer $L> \ell > 1$ and $\Bu\in (\DR^d)^N$ the cube $\bcube_L(\Bu)$ is $(E,\beta)$-CNR. Let $\bcube'\supset \bcube_{L+1}(\Bu)$,
$\By \in\bcube'\setminus \bcube_L(\Bu)$. Consider the lattice cubes $\bball_L(\Bu) \subset \bball_{L+1}(\Bu)$, and the function
$f: \bball_{L+1}(\Bu) \to \DR_+$ given by
$$
f: \, \Bx \mapsto \| \one_\By \BG_{\bcube'}(E) \one_\Bx\| .
$$
Let $\csS\subset\bball_{L - \ell - 1}(\Bu)$ be a (possibly empty) set such that any lattice cube
$\bball_\ell(\Bx)\subset \bball_{L - \ell - 1}(\Bu)\setminus \csS$ is $(E,\delta,m)$-NS.
If $0 < \beta < \delta \le 1$ and
$$
m\ell^\delta > 2L^\beta > L^\beta + \ln |\bball_L(\Bu)|,
$$
then  $f$ is $(\ell,q,\csS)$-dominated in $\bball_L(\Bu)$, with
$$
q =\eu^{-m' \ell^\delta}, \; \; m' := m - 2\ell^{-\delta}L^\beta >0.
$$
\ele

Now Lemma \ref{lem:good.NR.is.NS} can be proved essentially in the same way as \cite{CS13}*{Theorem 2.4.1},
with the help of Lemma \ref{lem:f.domin}. The role of the graph $\cG$ is played, of course, by the scatterers lattice
$\bcZ^N$.

\section{Proof of Lemma \ref{lem:prob.WI.S}}
\label{app:proof.WI.S}

\ble\label{lem:WITRONS.subexp}
Fix $\beta ,\delta\in (0,1]$, $m^*\geq 1$ and $E\in\DR$.
Suppose that a {\rm WI} ball $\bballN (\Bu,L_k)$
is $(E,\beta)$-NR and
satisfies the following two conditions:
\begin{align}
\label{eq:PNS.1}
\forall\, &\lam' \in\Sigma_{I^*}\left(\BH^{(N')}_{\bball'}\right) \;
\bball'' \text{ is } (E-\lam',\delta, m_{N'})-{\rm{NS}}
\\
\label{eq:PNS.2}
\forall\, &\lam'' \in\Sigma_{I^*}\left(\BH^{(N'')}_{\bball''}\right) \;
 \bball' \text{ is } (E-\lam'', \delta, m_{N''})-{\rm{NS}}.
\end{align}

If $L_0$ is large enough
then  $\bball^{(N)}(\Bu,L_k)$ is {\rm$(E, \delta, m_{N})${\rm-NS}}.
\ele

\proof
Recall that we assume the EVs of the operators appearing in our arguments to be numbered in increasing order.
We have the following identities:
\be\label{eq:Gni.proj.psi}
\BG_{\bcube_{L_k}(\Bu)}(E) = \sum_{a} \BP'_{\BPsi'_a}  \otimes \BG_{\bcube''}(E - E'_a)
= \sum_{a} \BG_{\bcube'}(E - E''_a) \otimes \BP''_{\BPsi''_a}
\ee
Further, the operator $\BG_{\bcube''}$ (as well as $\BG_{\bcube'}$) has compact resolvent, so $E'_a\uparrow +\infty$ as $a\to+\infty$.

By the second resolvent identity, for any energy $E$ which is not in the spectra of $\BHni_{\bcube}$ and $\BH_{\bcube}$,
we have for their respective resolvents $\BGni_\bcube(E)$ and $\BG_\bcube(E)$
$$
\BG_{\bcube} = \BGni_{\bcube} - \BGni_{\bcube} \BU_{\bcube',\bcube''} \BG_{\bcube}
$$
thus
$$
\bal
\| \chi_\By \BG \chi_\By  \| &\le \|\chi_\By \BGni \chi_\Bx \|  + \| \chi_\By \BGni \BU \BG \chi_\Bx \|
\\
& \le \|\chi_\By \BGni \chi_\Bx \|  + \| \BU_{\bcube',\bcube''}\| \| \BGni_{\bcube}\| \| \BG_{\bcube}\|.
\eal
$$

We start with the last term in the RHS.
Since $\BLam$ is weakly interactive, we have by
inequality \eqref{eq:Uone.norm.Ubcube.bcube} (cf. also  Lemma \ref{lem:WI.decomp})
$$
\| \BU_{\bcube',\bcube''} \| \le C \eu^{- (3NL_k)^\zeta} .
$$
The assumed $(E,\beta)$-NR property gives $\| \BG_{\bcube}\| \le \half \eu^{L_k^{\beta}}$; in terms of the spectrum
$\Sigma_\bcube$ of $\BH_{\bcube}$,
$\dist(E, \Sigma_\bcube)\ge 2\eu^{-L_k^{\beta}}$. The min-max principle implies for
the spectrum $\Sigmani_\bcube$ of $\BHni_{\bcube}$
\be\label{eq:perturbed.nonres.subexp}
\dist(E, \Sigmani_\bcube)\ge 2\eu^{-L_k^{\beta}} - \| \BU_{\bcube',\bcube''}\| \ge \eu^{-L_k^{\beta}},
\ee
so $\| \BGni_{\bcube}\| \le \eu^{L_k^{\beta}}$. Finally,
$$
\| \BU_{\bcube',\bcube''}\| \| \BGni_{\bcube}\| \| \BG_{\bcube}\| \le  C \eu^{- (3NL_k)^\zeta + 2 L_k^\beta}
\le \half \eu^{- L_k^\zeta }.
$$

It remains to assess the GF of the non-interacting Hamiltonian.
Denote
$$
a(\eta) = \max\{a: \, E'_a \le E_* + 2\eta\},
$$
then the Combes-Thomas estimate (cf. \cite{CT73,St01}) combined with the Weyl law implies that
$$
\sum_{a > a(\eta) } \dnorm{ \BP'_{\BPsi'_a} \otimes \BG_{\bcube''}(E - E'_a) }
\le \sum_{j=1}^{+\infty} L_k^{C} \eu^{ - (2\eta + j)L_k } \le \frac{1}{2} \eu^{ - \eta L_k}
\le \frac{1}{2} \eu^{ - 2m_N L_k^\delta}.
$$
It also follows from the Weyl law that $\card \{a: \, E''_a \le E_* + 2\eta\}\le L_k^{C'}$.
By assumption, for all $a\le a(\eta)$,
$$
\dnorm{ \BG_{\bcube''}(E - E'_a) }  \le \eu^{ - m_{N-1} L_k^\delta} \le \eu^{ - 2m_{N} L_k^\delta}.
$$
We conclude that
\be\label{eq:WI.NS.1}
\bal
\sum_{a}
\big\| \chi_{\By}  \BP'_{\BPsi'_a} \otimes \BG_{\bcube''}(E - E'_a) \chi_{\By} \big\|
  & \le \left( \sum_{a \le a(\eta)} + \sum_{a \le a(\eta)} \right) \BP'_{\BPsi'_a}  \otimes \BG_{\bcube''}(E - E'_a)
\\
& \le L_k^{C'} \, \eu^{ - 2m_{N} L_k^\delta} + \frac{1}{2} \eu^{ - 2m_N L_k^\delta}
\le \eu^{ - 2 m_N L_k^\delta}.
\eal
\ee
Similarly,
\be\label{eq:WI.NS.2}
\bal
\sum_{a}
\big\| \chi_{\By}  \BG_{\bcube'}(E - E''_a)\otimes \BP''_{\BPsi''_a} \chi_{\By} \big\|
  & \le \eu^{ - 2 m_N L_k^\delta}.
\eal
\ee
Taking the sum over all $\By\in \pt^{-} \bball_{L_k}(\Bu)$, falling into one of the two categories
\eqref{eq:WI.NS.1}--\eqref{eq:WI.NS.2}, we obtain for $L_0$ large enough
$$
\dnorm{ \BG_{\bcube_{L_k}(\Bu)}(E) } \le \Const L_k^{Nd} \eu^{ -2 m_N L_k^\delta}
\le \eu^{ -m_N L_k^\delta},
$$
which proves the assertion of the lemma.
\qedhere

\proof[Proof of Lemma \ref{lem:prob.WI.S}]

Denote by $\cS$ the event in the LHS of \eqref{eq:lem.prob.WI.S.1}. Let $\bcube=\bcube^{(N)}(\Bu,L_k)$ and
consider the canonical factorization $\bcube=\bcube'\times\bcube''$.
We have
\be\label{eq:proof.lem.WI.T.a}
\bal
\pr{\cS } & < \pr{\text{ $\bcube$  is not  \ENR }}
\\
&
+ \pr{\text{ $\bcube$  is  \ENR and $E,\delta,m_N$-S }}.
\eal
\ee
By Theorem \ref{thm:W1}, the first term in the RHS is bounded by
$\eu^{-L_{k+1}^{\beta}}< \frac{1}{3} \eu^{-\frac{3}{2}\nu_N L_{k+1}^{\kappa}}$,
since $\kappa < \beta$, so we focus on the second summand.

Let $\Sigma' =\Sigma\big(\BH^{(N')}_{\bball'}\big)\cap I^*$,
$\Sigma'' =\Sigma\big(\BH^{(N'')}_{\bball''}\big)\cap I^*$,
and consider the events
$$
\bal
\cS' &= \{\om:\, \exists\, \lam' \in \Sigma', \;
\bball'' \text{ is } (E-\lam',\delta, m_{N'})-{\rm{NS}} \} ,
\\
\cS'' &= \{\om:\,  \exists\, \lam'' \in\Sigma'', \;
 \bball' \text{ is } (E-\lam'', \delta, m_{N''})-{\rm{NS}} \}.
\eal
$$
Since $\bcube$ is WI, we have that
$\Pi \bcube' \cap \Pi \bcube'' = \varnothing$, $\BH_{\bcube''}(\om)$ is independent of
the sigma-algebra $\fF'$ generated by the random scatterers affecting $\bcube'$, while
$\BH_{\bcube'}(\om)$ is $\fF'$-measu\-rable, and so are all the EVs $\lam'\in\Sigma'$.

Further, by non-negativity of $\BH'$, if $E \le E^*$, then $E - \lam' \le E^*$ for all $\lam' \in \Sigma'$.

Replacing the quantity $E-\lam'$, rendered nonrandom by conditioning on $\fF'$, with a new nonrandom
parameter $E'\le E^*$, we have, by induction in $1 \le n \le N-1$, and with $\nu_{N''}\ge \nu_{N-1}$,
\be\label{eq:prob.cS.1}
\bal
\pr{\cS'} &= \esm{ \pr{\csS' \,|\, \fF''} } \le \sup_{ E' \le E^*} \pr{ \bcube'' \text{ is $(E',m)$-S}}
\\
& \le |\bcube''|\, \eu^{ - \nu_{N''} L_k^\kappa} \le |\bcube''|\, \eu^{ - 2\nu_N L_k^\kappa}
\le \third \eu^{ - \frac{3}{2}\nu_N L_k^\kappa} .
\eal
\ee
Similarly,
\be\label{eq:prob.cS.2}
\bal
\pr{\cS''}  \le \third \eu^{ - \frac{3}{2}\nu_N L_k^\kappa} .
\eal
\ee
Collecting \eqref{eq:proof.lem.WI.T.a}--\eqref{eq:prob.cS.2},
the assertion \eqref{eq:lem.prob.WI.S.1} follows.

For the second assertion \eqref{eq:lem.prob.WI.S.2}, it suffices to apply a polynomial bound on the number of
cubes of size $L_k$ with centers on the lattice $\bcZ^N$ in a cube of radius $L_{k+1}$.
\qedhere


\end{document}